\documentclass[preprint,12pt]{elsarticle}
\usepackage{amsmath,amssymb,amsthm}
\pdfoutput=1
\usepackage{mathrsfs}
\usepackage{subfig}
\usepackage{courier}
\usepackage{graphicx}
\usepackage{pst-node}
\usepackage{algorithm}
\usepackage{pdflscape}
\usepackage{algorithmicx}
\usepackage{algpseudocode}
\usepackage[abs]{overpic}
\usepackage{tikz}
\usepackage{url}
\usepackage{color,soul}
\usepackage{textcomp}
\usepackage{graphicx}
\usepackage{caption}
\usepackage{tikz}
\usepackage{hyperref}
\usepackage{upgreek}

\newtheorem{remark}{Remark}
\biboptions{comma,sort&compress}
\usepackage{titlesec}

\usepackage{hyperref}
\usepackage{mathtools}

\titleclass{\subsubsubsection}{straight}[\subsection]

\newcounter{subsubsubsection}
\renewcommand\thesubsubsubsection{\thesubsubsection.\arabic{subsubsubsection}}

\titleformat{\subsubsubsection}
  {\normalfont\normalsize}{\thesubsubsubsection}{1em}{}
\titlespacing*{\subsubsubsection}
{0pt}{3.25ex plus 1ex minus .2ex}{1.5ex plus .2ex}

\makeatletter
\renewcommand\paragraph{\@startsection{paragraph}{5}{\z@}%
  {3.25ex \@plus1ex \@minus.2ex}%
  {-1em}%
  {\normalfont\normalsize\bfseries}}
\renewcommand\subparagraph{\@startsection{subparagraph}{6}{\parindent}%
  {3.25ex \@plus1ex \@minus .2ex}%
  {-1em}%
  {\normalfont\normalsize\bfseries}}
\def\toclevel@subsubsubsection{4}
\def\toclevel@paragraph{5}
\def\toclevel@paragraph{6}
\def\l@subsubsubsection{\@dottedtocline{4}{7em}{4em}}
\def\l@paragraph{\@dottedtocline{5}{10em}{5em}}
\def\l@subparagraph{\@dottedtocline{6}{14em}{6em}}
\makeatother

\begin{document}
\newcommand{\comment}[1]{\textcolor{red}{#1}}
\newcommand{\jump}[1]{\text{\textlbrackdbl}#1\text{\textrbrackdbl}}
\makeatletter
\newcommand*{\rom}[1]{\expandafter\@slowromancap\romannumeral #1@}
\makeatother
\begin{frontmatter}
\title{An energy-stable convex splitting for \\ the phase-field crystal equation}
\author[numpor,MSE]{P.~Vignal\corref{cor1}\fnref{fn1}}
\author[numpor,conicet]{L.~Dalcin}
\author[numpor]{D.L.~Brown}
\author[numpor]{N.~Collier}
\author[numpor,kaust]{V.M.~Calo}
\address[numpor]{Center for Numerical Porous Media (NumPor)\\
King Abdullah University of Science and Technology (KAUST)\\Thuwal, Saudi Arabia}
\address[MSE]{Materials Science and Engineering (MSE)\\King Abdullah University of Science and Technology (KAUST)\\Thuwal, Saudi Arabia}
\address[conicet]{Consejo Nacional de Investigaciones Cient\'\i{}ficas y T\'ecnicas (CONICET)\\Santa Fe, Argentina}
\address[kaust]{Applied Mathematics and Computational Science (AMCS)\\Earth Science and Engineering (ErSE)\\
King Abdullah University of Science and Technology (KAUST)\\Thuwal, Saudi Arabia}
\cortext[cor1]{Corresponding author}
\fntext[fn1]{Tel.: $966~012~808~0391$, E-mail address: philippe.vignal@kaust.edu.sa.}
\begin{abstract}
The phase-field crystal equation, a parabolic, sixth-order and nonlinear partial differential equation, has generated considerable interest as a possible solution to problems arising in molecular dynamics. Nonetheless, solving this equation is not a trivial task, as energy dissipation and mass conservation need to be verified for the numerical solution to be valid. This work addresses these issues, and proposes a novel algorithm that guarantees mass conservation, unconditional energy stability and second-order accuracy in time. Numerical results validating our proofs are presented, and two and three dimensional simulations involving crystal growth are shown, highlighting the robustness of the method.
\end{abstract}
\end{frontmatter}
\section{Introduction}\label{s:intro}
While the tight connection between material processing, structure and properties has been known for years, a microstructural model capable of accounting for the atomic scale features affecting the macroscale properties of a material has not yet been established. Progress has nonetheless been made in this direction, and this work tackles one of the solution strategies that has recently been proposed through the phase-field crystal equation~(PFC). Developed as an extension to the phase-field formalism in which the fields take spatially uniform values at equilibrium~\cite{Elder0,Elder1}, the free energy functional in the case of the PFC equation is minimized by periodic states. These periodic minima allow this particular phase-field model to represent crystalline lattices in two and three dimensions~\cite{Elder2010,Jaatinenext}, and more importantly, to capture the interaction of defects that arise at the atomic scale without the use of additional fields, as is done in standard phase-field equations~\cite{Rappel}. This model has also been shown to successfully cross time scales~\cite{Elder}, thanks in part to the phase-field variable that describes a coarse grained temporal average (the number density of atoms). This difference in time scale with molecular dynamics, along with the periodic density states that naturally give rise to elasticity, multiple crystal orientations, and the nucleation and motion of dislocations, are some of the reasons why this tool is being considered for quantitative modeling~\cite{Provatas,Asadi}.

Several challenges are unfortunately faced while simulating the PFC numerically. It is a sixth-order, nonlinear, partial differential equation, where the solution should lead to a time-decreasing free energy functional. Recent work on this topic includes~\cite{weakenergy,Hu2009,elsey,Gomez2012,GomezH,Tegze2009}. Inspired by the work presented for the Cahn--Hilliard equation in the context of tumor-growth~\cite{kris}, we developed a formulation capable of conserving mass, guaranteeing discrete energy stability while having second-order temporal accuracy. The numerical scheme achieves this through a convex splitting of the nonlinearity present in the equation, along with the addition of a stabilization term, while using a mixed form that segregates the partial differential equation into a system of three second order equations. This is similar in a sense to what was done in~\cite{Gomez2012}, where a mixed form is also used, but has the added advantage that the well-posedness of the variational form does not require globally $\mathcal{C}^1$-continuous basis functions. This presents an advantage in terms of computational cost~\cite{Collier0,Collier1,Collier2} as linear, $\mathcal{C}^0$ finite elements can be used.

We provide mathematical proofs for mass conservation, energy stability and second-order accuracy, properties that the algorithm possesses, along with two-dimensional numerical evidence that corroborates our findings. We also present three dimensional results that showcase the effectiveness and robustness of our algorithm. The paper is structured as follows: In section~\ref{s:pfc}, we describe the phase field crystal equation. In section~\ref{s:numfor}, we present our numerical scheme. Section~\ref{s:cost} presents numerical examples dealing with crystal growth in a supercooled liquid. We give concluding remarks in section~\ref{s:conc}.


\section{Phase-field crystal model}\label{s:pfc}By using a free energy functional that is minimized by periodic density fields, the phase-field crystal equation is capable of representing crystalline lattices~\cite{Elder0}, and more importantly, capturing the interaction between material defects implicitly. The model is characterised by a conserved field related to the atomic number density, that is spatially periodic in the solid phase and constant in the liquid phase. It has been related successfully to other continuum field theories such as density-functional theory~\cite{Elder,Ramakrishnan}. This work will show examples related to crystalline growth, as the PFC equation has found much of its success in modelling microstructural evolution~\cite{Elder1,Backofen,Backofen1,Elder,gyula}, while it has also been used to model other physical phenomena such as foam dynamics~\cite{Guttenberg}, glass formation~\cite{Berry}, liquid crystals~\cite{Lowen}, elasticity~\cite{Elder0} and in the estimation of material properties~\cite{Pisutha}. 

Experimental and computational results can differ significantly, but work is nonetheless being done to reduce the mismatch~\cite{Wu_bcc, Wu_fcc, Pisutha,Jaatinen}. The model that is considered in this work can be improved by increasing the number of critical wavelengths one considers in the free energy functional at the expense of computational cost, as the partial differential equation becomes harder to solve~\cite{Asadi,Wu_fcc}. Also, molecular dynamics in a multi-scale setting can be used to estimate some of the parameters going into the phase-field crystal equation~\cite{Asadi1}, and inverse formulations of the problem could be considered to validate the calculations~\cite{Wu}. Hopefully, these multi-scale approaches will allow for more complete studies on polycrystalline growth using the PFC equation, such as the ones presented in~\cite{Jamshidian, Thamburaja} in the setting of phase-field modeling.
\subsection{Model formulation}
\indent The phase-field crystal equation was developed to study the evolution of microstructures, at atomic length scales and diffusive time scales, by considering a conservative description of the Rayleigh-B\'enard convection problem~\cite{Elder2010}. The order parameter $\phi$ represents an atomistic density field in the model, which is periodic in the solid state and uniform in the liquid one. The free energy functional for the phase-field crystal equation in its dimensionless form is given by~\cite{Elder1,Gomez2012,Jaatinenext}
\begin{align}\label{FEright}
\mathcal{F}[\phi(\mathbf{x})]=\int_{\Omega}\left[\Psi(\phi)+\frac{1}{2}\left(\phi^2-2|\nabla\phi|^2+(\Delta\phi)^2\right) \right]d\mathbf{\Omega},
\end{align}
where $\Omega \in \mathbb{R}^d$ represents an arbitrary open domain, with $d=2$ or $3$, and $\Psi\left(\phi\right)=-\dfrac{\epsilon}{2}\phi^2+\dfrac{1}{4}\phi^4$. The parameter $\epsilon$ represents a critical transition variable, which in the case of crystal growth is associated to the degree of undercooling: the larger its value, the larger the undercooling is. The free energy functional presented in equation~\eqref{FEright} is then minimized to achieve thermodynamical stability. To enforce this mathematically, one solves the Euler--Lagrange equation for the free energy, and takes its variational derivative with respect to $\phi$. The variational derivative is given by
\begin{align}\label{variational}
\frac{\delta \mathcal{F}}{\delta \phi}&=\frac{\partial \mathcal{F}}{\partial \phi}-\nabla \cdot \frac{\partial \mathcal{F}}{\partial \nabla \phi}+\Delta \frac{\partial \mathcal{F} }{\partial \Delta\phi} \nonumber\\
&=(1+\Delta)^2\phi+\Psi'(\phi),
\end{align}
where $\nabla \cdot$, $\nabla$ and $\Delta$ denote the divergence, gradient and Laplacian operators, respectively, and $\Psi'(\phi)=-\epsilon\phi +\phi^3$ with $(1+\Delta)^2=1+2\Delta+\Delta\Delta$. The partial differential equation, considering that the atomistic density field is a conserved quantity~\cite{Elder1}, is then formulated as 
\begin{align}\label{PDE}
\frac{\partial{\phi}}{\partial{t}}=\nabla \cdot \left(M\nabla\frac{\delta \mathcal{F}}{\delta{\phi}} \right),
\end{align}
where $\phi\equiv\phi\left(\textbf{x},t\right)$ represents the phase field, $\mathbf{x}$ and $t$ represent space and time, respectively, $M$ is the mobility, and $\mathcal{F}$ is the free energy functional of the system. The partial differential equation, after substituting equation (\ref{variational}) into (\ref{PDE}), becomes
\begin{align*}\label{6PDE}
\frac{\partial \phi}{\partial t}&=\nabla\cdot\nabla\left[\left(1+\Delta\right)^2\phi+\Psi'(\phi)\right] \nonumber\\
&=\Delta\left[\left(1+\Delta\right)^2\phi+\Psi'(\phi)\right],
\end{align*}
where the mobility $M$ is assumed equal to a constant of value one.
\subsection{Phase-field crystal equation: strong form}
The problem is stated as follows: over the spatial domain $\Omega$ and the time interval $]0,T[$, given  $\phi_0:{\Omega}\longmapsto\mathbb{R}$, find $\phi:\Omega\times [0,T]\longmapsto \mathbb{R}$ such that
\begin{equation}\label{PDE1}
\begin{cases}
\dfrac{\partial \phi}{\partial t} = \Delta \left[\left(1+\Delta\right)^2\phi+\Psi'(\phi)\right] \  &\text{on}\ \Omega \times ]0,T], \\
\phi(\mathbf{x},0) = \phi_0(\mathbf{x})\quad &\text{on}\ \Omega,
\end{cases}
\end{equation}
where $\phi_0(\mathbf{x})$ represents a function that approximates a crystalline nucleus, and periodic boundary conditions are imposed in all directions. We discuss the choices made to handle initial conditions in section~\ref{s:cost}.

\section{Stable time discretization for the phase-field crystal equation}\label{s:numfor}
The phase-field crystal equation is a sixth-order, parabolic partial differential equation. If an explicit time-stepping scheme were employed to solve it, a time step size $\Delta t$ on the order of the sixth power of the grid size would be required. This restriction has motivated research in implicit algorithms~\cite{Gomez2012,Hu2009,weakenergy,elsey} and adaptive algorithms~\cite{Zhang3}. On top of this, some properties need to be guaranteed while solving the equation, such as mass conservation, defined as
\begin{align}\label{eq:mc}
\int_\Omega\left(\frac{\partial{\phi}}{\partial t}\right)d\mathbf{\Omega}=0
\end{align}
due to the fact that density is conserved, as well as strong energy stability~\cite{weakenergy}, expressed as
\begin{align}\label{eq:femin}
\mathcal{F}\left[\phi\left(t_{n+1}\right)\right]\leq \mathcal{F}\left[\phi\left(t_n\right)\right] \quad \forall n=1, 2, ..., N,
\end{align}
which translates to having the free energy be monotonically decreasing. In this work, we develop an algorithm that extends the ideas presented in~\cite{kris,elsey}, guarantees the properties presented in equations~\eqref{eq:mc} and~\eqref{eq:femin}, while achieving second-order accuracy in time. The discretization in space is done using isogeometric analysis~(IGA), a finite element method where NURBS are used as basis functions~\cite{Cottrell2009}. The method not only allows to control the spatial resolution of the mesh~($h$-refinement) and the polynomial degree of the basis~($p$-refinement), but also to increase their global continuity~($k$-refinement). Isogeometric analysis has successfully been applied to phase-field modeling~\cite{Liu, calo,Gomez,GomezH,Gomez2012,borden,iccs_vignal,serbia_vignal}. The PFC model, being a nonlinear, sixth-order in space, first-order in time partial differential equation, allows for many choices in terms of discretizations and time stepping schemes. High-order, globally continuous basis functions can be easily generated within the IGA framework, reason why it allows for the straightforward discretization of high-order partial differential equations. Alternatively, mixed formulations can be employed so as to reduce the continuity requirements down to standard $\mathcal{C}^0$ spaces used in traditional finite element methods. This work makes use of a mixed form, where the system that is solved involves a coupled system of three second-order equations.
\subsection{Mixed form $2+2+2$: triple second-order split}
Equation (\ref{PDE1}) can be written as a system that consists of three coupled second-order equations, given by
\begin{subequations}\label{triple}
\begin{align}
\frac{\partial \phi}{\partial t}&=\Delta \sigma&  \text{\space \space  in } \Omega\times]0,T],\label{triple1}\\
\sigma&=\left(1+\Delta\right)\theta + \Psi'\left(\phi\right) &\text{\space \space  in } \Omega\times]0,T],\label{triple2}\\
\theta&= \left(1+\Delta\right)\phi & \text{\space \space  in } \Omega\times]0,T].\label{triple3}
\end{align}
\end{subequations}
\subsubsection{Weak form}
Let us denote by $\mathcal{V}_1$ a functional space, which is a subset of~$\mathcal{H}^1$, where~$\mathcal{H}^1$ is the Sobolev space of square integrable functions with square integrable first derivatives. Assuming periodic boundary conditions in all directions, a weak form can be derived by multiplying~(\ref{triple1}) to~(\ref{triple3}) by test functions~$q, s, w \in \mathcal{V}_1$, respectively, and integrating the equations by parts. The variational problem can then be defined as that of finding~$\phi, \theta, \sigma \in \mathcal{V}_1$ such that for all~$q, s, w \in \mathcal{V}_1$
\begin{align*}
0&=\left(q,\dot{\phi}\right)_{\Omega} + (\nabla q,\nabla \sigma)_{\Omega}\nonumber\\
&+\left(s,\sigma-\Psi'\left(\phi\right)-\theta\right)_{\Omega}+\left(\nabla s,\nabla \theta\right)_{\Omega}\nonumber\\
&+\left(w,\theta-\phi\right)_{\Omega}+\left(\nabla w,\nabla \phi\right)_{\Omega},
\end{align*}
where the dependence of $\phi$ on space and time is not explicitly stated, the $\mathcal{L}^2$ inner product over the domain $\Omega$ is indicated by $(.,.)_{\Omega}$ and $\dot{\phi}:=\dfrac{\partial {\phi}}{\partial t}$.
\subsubsection{Semi-discrete formulation}
Splitting the equation with the help of the auxiliary variables $\sigma$ and $\theta$ allows us to use $C^0$ finite elements, as only $\mathcal{H}^1$-conforming spaces are needed. We let $\mathcal{V}_1^h\subset \mathcal{V}_1$ denote the finite dimensional functional space spanned by these $C^0$~B-spline basis functions in two or three spatial dimensions. The problem is then stated as follows: find~$\phi^h, \theta^h, \sigma^h \in \mathcal{V}_1^h$ such that for all~$q^h, s^h, w^h \in \mathcal{V}_1^h$
\begin{align}\label{222}
0&=\left(q^h,\dot{\phi^h}\right)_{\Omega} + (\nabla q^h,\nabla \sigma^h)_{\Omega}\nonumber\\
&+\left(s^h,\sigma^h-\Psi'\left(\phi^h\right)-\theta^h\right)_{\Omega}+\left(\nabla s^h,\nabla \theta^h\right)_{\Omega}\nonumber\\
&+\left(w^h,\theta^h-\phi^h\right)_{\Omega}+\left(\nabla w^h,\nabla \phi^h\right)_{\Omega},
\end{align}
where the weighting functions $q^h$, $s^h$ and $w^h$, and trial solutions  $\sigma^h$, $\theta^h$ and $\phi^h$ can be defined as
\begin{align*}
q^h=\sum_{A=1}^{n_b}q_AN_A, &\quad s^h=\sum_{A=1}^{n_b}s_AN_A, \quad w^h=\sum_{A=1}^{n_b}w_AN_A, \\
\sigma^h=\sum_{A=1}^{n_b}\sigma_AN_A, &\quad \theta^h=\sum_{A=1}^{n_b}\theta_AN_A, \quad \phi^h=\sum_{A=1}^{n_b}\phi_AN_A,
\end{align*}
where the B-spline basis functions $N_A$ define the discrete space $\mathcal{V}^h_1$ of dimension $n_b$ and the coefficients $q_A$, $s_A$, $w_A$, $\sigma_A$, $\theta_A$ and $\phi_A$ represent the control variables.
\subsubsection{Time discretization}\label{subset:time}
The time discretization proposed in this work adapts what was done in~\cite{kris} for the Cahn--Hilliard equation, to the formulation presented in equation~\eqref{222} for the phase-field-crystal equation. To do this, the nonlinear term $\Psi(\phi)=\dfrac{\phi^4}{4}-\dfrac{\epsilon\phi^2}{2}$ is split as
\begin{align*}
\Psi(\phi)=\Psi_c(\phi)-\Psi_e(\phi),
\end{align*}
where $\Psi_c(\phi)=\dfrac{\phi^4}{4}$ and $\Psi_e(\phi)=\dfrac{\epsilon\phi^2}{2}$. Both of these functions are convex, which allows us to discretize the nonlinearity in time using a convex-implicit, concave-explicit treatment, giving the following fully discrete system %
\begin{align}
0&=\left(q^h,\frac{\jump{ \phi_{n}^h}}{\Delta t}\right)_{\Omega} + \left(\nabla q^h, \nabla \sigma^h\right)_{\Omega}\nonumber\\
&+\left(s^h,\sigma^h-\theta^h-\left(\Psi_{c}'\left(\phi^h_{n+1}\right)-\Psi_{c}''\left(\phi^h_{n+1}\right)\frac{\jump{\phi_n^h}}{2}\right)\right)_{\Omega}\nonumber\\
&+\left(s^h,\left(\Psi_{e}'\left(\phi^h_{n}\right)+\Psi_{e}''\left(\phi^h_{n}\right)\frac{\jump{\phi_n^h}}{2}\right)\right)_{\Omega}\nonumber\\
&+\left(\nabla s^h,\nabla \theta^h -\alpha_n\Delta t \nabla \jump{\phi_n^h}\right)_{\Omega}\nonumber\\
&+\left(w^h,\theta^h-\{\phi^h_n\} \right)_{\Omega} + \left(\nabla w^h,\nabla \{\phi_n^h\} \right)_{\Omega},\label{eq:triple1}
\end{align}
where
\begin{itemize}
\item $\jump{\phi_n^h}=\phi^h_{n+1}-\phi^h_{n}$,
\item $\{\phi^h_n\}=\dfrac{1}{2}\left(\phi^h_{n+1}+\phi^h_n\right)$,
\item $\Psi'_c({\phi_{n+1}^h})=\left(\phi^h_{n+1}\right)^3$,
\item $\Psi''_c({\phi_{n+1}^h})=3\left(\phi^h_{n+1}\right)^2$,
\item $\Psi'_e({\phi_{n}^h})=\epsilon\phi^h_n$,
\item $\Psi''_e({\phi_{n}^h})=\epsilon$,
\end{itemize}
and the stabilization parameter $\alpha_n$ needs to comply with
\begin{align*}
\alpha_n\geq\dfrac{\left(\text{sup}\left(\Psi_c''\left(\phi^h_{n+1}\right)+\Psi_e''\left(\phi^h_{n}\right)\right)\right)^2}{16}=\dfrac{\left(\text{sup}\left(3\left(\phi^h_{n+1}\right)^2+\epsilon\right)\right)^2}{16}
\end{align*}
\subsubsection{Properties of the numerical scheme}
The discretization presented in section~\ref{subset:time} guarantees mass conservation, is second-order accurate in time, and possesses energy stability by construction.
\subsubsubsection{Mass conservation}
Mass conservation can be verified by taking equation~(\ref{eq:triple1}), and letting the test function $q^h$ be equal to one while having $s^h=w^h=0$, such that
\begin{align*}
0=&\left(1,\dfrac{\jump{\phi^h_n}}{\Delta t}\right)_{\Omega}+\left(0,\nabla \sigma^h\right)=\int_{\Omega}\dfrac{\jump{\phi_n^h}}{\Delta t}d\mathbf{\Omega},
\end{align*} 
which implies that mass is conserved at the discrete time levels, that is
\begin{align*}
\int_{\Omega}\phi_{n+1}d\mathbf{\Omega} = \int_{\Omega}\phi_{n}d\mathbf{\Omega}.
\end{align*}
\subsubsubsection{Second-order accuracy in time}
A bound on the local truncation error can be obtained by comparing our method to the Crank-Nicolson scheme, a well-known second-order accurate time-stepping algorithm. If we do not spatially discretize~\eqref{PDE1}, but instead apply the Crank-Nicolson scheme to it, we obtain 
\begin{align*}
\dfrac{\jump{ \phi_n }  }{\Delta t} = \Delta \left[\left(1+\Delta\right)^2\{\phi_n\}+\Psi'(\{\phi_n\})\right].
\end{align*}
Substituting the discrete time solution $\{\phi_n\}$ by the time-continuous solution~$\phi(t_n)$ into the above equation gives rise to the local truncation error. Indeed, we have 
\begin{align}
\dfrac{\jump{ \phi(t_n) }  }{\Delta t} = \Delta \left[\left(1+\Delta\right)^2\{\phi(t_n)\}+\Psi'(\{\phi(t_n)\})\right]+\tau(t_n),
\end{align}
where~$\tau(t_n)$ represents the global truncation error. It can be showh, using Taylor series,  that such a scheme will give a bound $\tau(t_n)\leq C \Delta t^2$, as was done in \cite{GomezH} in a similar context.

To prove second-order accuracy in time for our scheme, we compute the next time-step approximation via the scheme applied to the exact solution and compare the result to Taylor expansions. 
A similar procedure was performed in  \cite{kris} in the context of Cahn--Hilliard equations. By looking at only the time discretization part of \eqref{eq:triple1}, and reorganizing the splitting into one equation, we have that
\begin{align}
\phi_{n+1}=\phi(t_{n})+\Delta t\Delta \Bigg(    &(1+\Delta)^2\phi(\{t\})\nonumber\\
&+\Psi_{c}'(\phi(t_{n+1}))-\Psi_{e}'(\phi(t_{n}))\nonumber\\\label{approx}
&-\frac{1}{2}\jump{\phi(t)}\Psi_{c}^{''}(\phi(t_{n+1}))-\frac{1}{2}\jump{\phi(t)}\Psi_{e}^{''}(\phi(t_{n}))\nonumber\\
&\qquad \qquad \qquad \qquad \qquad \qquad -\alpha_{n} \Delta t \Delta \jump{\phi(t)}\Bigg),
\end{align}
where $\phi(\{t\})$ is defined as the Crank-Nicolson (mid-point rule) approximation
\begin{align*}
\phi(\{t\})=\phi\Bigg(\frac{t_{n+1}+t_{n}}{2}\Bigg)=\frac{\phi(t_{n+1})+\phi(t_{n})}{2} +\mathcal{O}(\Delta t^2).
\end{align*}
We  expand $\Psi_{c}'\left(\phi\left(t_{n+1}\right)\right)$ such that
\begin{align*}
\Psi_{c}'(\phi(t_{n+1}))&=\Psi_{c}'(\phi(\{t\}))-\Psi_{c}''(\phi(t_{n+1}))\big(\phi(\{t\})-\phi(t_{n+1})\big)+\mathcal{O}(\Delta t^2)\nonumber \\
&=\Psi_{c}'(\phi(\{t\}))+\frac{\jump{\phi({t})}}{2}\Psi_{c}''(\phi(t_{n+1}))+\mathcal{O}(\Delta t^2).
\end{align*}
Thus, 
\begin{align}\label{Psi1}
\Psi_{c}'(\{\phi\})=\Psi_{c}'(\phi(t_{n+1}))-\frac{\jump{\phi(t)}}{2}\Psi_{c}''(\phi(t_{n+1}))+\mathcal{O}(\Delta t^2).
\end{align}
Similarly, we have for the explicit part
\begin{align}\label{Psi2}
\Psi_{e}'(\{\phi\})=\Psi_{e}'(\phi(t_{n}))+\frac{\jump{\phi(t)}}{2}\Psi_{e}''(\phi(t_{n}))+\mathcal{O}(\Delta t^2).
\end{align}
The stabilization term is of order $\mathcal{O}(\Delta t^2)$ and  can be written as 
\begin{align}
\alpha_{n} \Delta t \Delta \jump{\phi(t)}&=\alpha_{n} (\Delta t )^2\Delta\Bigg( \frac{\jump{\phi(t)}}{\Delta t}\Bigg)\nonumber \\
\label{alphaterm}
&=\alpha_{n} (\Delta t )^2\Delta\Bigg( \frac{\partial {\phi}}{\partial  t}+\mathcal{O}\big(\Delta t\big)\Bigg)=\mathcal{O}(\Delta t^2).
\end{align}
Using \eqref{Psi1}-\eqref{alphaterm}, and substituting them into \eqref{approx}, we obtain 
\begin{align}\label{approx2}
\phi_{n+1}=\phi(t_{n})+\Delta t\Delta \Bigg(    &(1+\Delta)^2\phi(\{t\})\nonumber\\
&+\Psi_{c}'(\{\phi\})-\Psi_{e}'(\{\phi\})+\mathcal{O}(\Delta t^2)\Bigg).
\end{align}
Alternatively, by Taylor expansion of the solution, we have 
\begin{align*}
\phi(\{t\})&=\phi(t_{n+1})-\frac{\Delta t}{2}\phi'(\{t\})-\frac{1}{2}\Bigg(\frac{\Delta t}{2}\Bigg)\phi''(\{t\})+\mathcal{O}(\Delta t^3),\\
\phi(\{t\})&=\phi(t_{n})+\frac{\Delta t}{2}\phi'(\{t\})-\frac{1}{2}\Bigg(\frac{\Delta t}{2}\Bigg)\phi''(\{t\})+\mathcal{O}(\Delta t^3).
\end{align*}
Taking the difference of the above two equations and using \eqref{PDE1} yields 
\begin{align*}
\phi(t_{n+1})-\phi(t_{n})&=\Delta t \frac{\partial \phi(\{t\} )}{\partial t}+\mathcal{O}(\Delta t^3)\nonumber\\
&=\Delta t\Delta \left((1+\Delta)^2\phi(\{t\})+\Psi_{c}'(\{\phi\})-\Psi_{e}'(\{\phi\})+\mathcal{O}(\Delta t^3)\right).
\end{align*}
Finally, taking the difference of the above expression with \eqref{approx2}, we obtain the local truncation error
\begin{align*}
\phi(t_{n+1})-\phi_{n+1}=\mathcal{O}(\Delta t^3).
\end{align*}
Thus, using the fact that the global truncation error $\tau(t_n)$ loses an order of $\Delta t$, the scheme is second-order accurate in time.
\subsubsubsection{Energy stability}
To prove energy stability, we first consider the time-discrete  form of the scheme, given by
\begin{align}
\jump{\phi_n}&=\Delta t \Delta\tilde{\sigma},\\
\tilde{\sigma}&=\left(1+\Delta\right)\tilde{\theta}-\alpha_n\Delta t \Delta\jump{\phi_n}\nonumber\\
&+\Psi'_{c}\left(\phi_{n+1}\right)-\dfrac{1}{2}\jump{\phi_n}\Psi_{c}''\left(\phi_{n+1}\right)-\Psi_{e}'\left(\phi_{n}\right)-\dfrac{1}{2}\jump{\phi_n}\Psi''_{e}\left(\phi_{n}\right),\label{eq:sf2222}\\
\tilde{\theta}&=\left(1+\Delta\right)\{\phi_n\}.\label{eq:sf2223}
\end{align}
Considering that for any smooth function $\Psi$ we have
\begin{align*}
\jump{\Psi}&=\Psi'(\phi_{n})\jump{\phi}+\Psi''(\xi_1(\phi_{n+1},\phi_n))\frac{\jump{\phi}^2}{2}\\
&=\Psi'(\phi_{n+1})\jump{\phi}-\Psi''(\xi_2(\phi_{n+1},\phi_n))\frac{\jump{\phi}^2}{2},
\end{align*}
for  some $\xi_1(\phi_{n+1},\phi_n)$  in between $\phi_{n}$ and $\phi_{n+1}$, similarly for $\xi_2(\phi_{n+1},\phi_n)$. 
The above formula is the exact Taylor series with remainder term and no additional terms are required in the expansion. 

Applying these expansions to our particular form of the nonlinearity, by Taylor's theorem, for some $\xi_{c},\xi_{e}$ in between $\phi_{n}$ and $\phi_{n+1}$, we have that
\begin{align*}
\jump{\Psi}&=\jump{\Psi_{c}}-\jump{\Psi_{e}}\nonumber\\
&= \Psi'_{c}(\phi_{n+1})\jump{\phi}-\Psi''_{c}(\xi_{c})\frac{\jump{\phi}^2}{2}-\Psi'_{e}(\phi_{n})\jump{\phi}
-\Psi''_{e}(\xi_{e})\frac{\jump{\phi}^2}{2}.
\end{align*}
Since $\Psi_{c}$ and $\Psi_{e}$ are globally convex, we have that $\Psi''_{c}, \Psi''_{e} \geq 0$. Observe that
\begin{align}\label{convexity1}
\jump{\Psi} &= \left(\Psi'_{c}(\phi_{n+1})-\Psi'_{e}(\phi_{n})\right)\jump{\phi}-\left(\Psi''_{c}(\xi_{c})
+\Psi''_{e}(\xi_{e})\right)\frac{\jump{\phi}^2}{2}\nonumber\\
&\leq \left(\Psi'_{c}(\phi_{n+1})-\Psi'_{e}(\phi_{n})\right)\jump{\phi}.
\end{align}
Here, we use the fact that the second derivatives are non-negative and the overall sign of the second derivative terms is negative.
Recalling equation~(\ref{FEright}), we have that the free energy is given by
\begin{align*}
\mathcal{F}[\phi(\mathbf{x})]=\int_{\Omega}\left[\Psi(\phi)+\frac{1}{2}\left(\phi^2-2|\nabla\phi|^2+(\Delta\phi)^2\right) \right]d\mathbf{\Omega},
\end{align*}
with which we can write, given equation~\eqref{convexity1}, that
\begin{align*}
\jump{\mathcal{F}[\phi(\mathbf{x})]}&=\jump{\mathcal{F}}\nonumber\\
&=\int_{\Omega}\left(\jump{\Psi\left(\phi\right)} + \frac{1}{2}\jump{\left(\phi\right)^2-2|\nabla\phi|^2+(\Delta\phi)^2}\right)d\mathbf{\Omega}\nonumber\\
&\leq\int_{\Omega}\left(\Psi'_{c}\left(\phi_{n+1}\right)-\Psi'_{e}\left(\phi_{n}\right)\right)\jump{\phi}d\mathbf{\Omega}\nonumber\\
\qquad &+\dfrac{1}{2}\int_{\Omega} \jump{\left(\left(\phi\right)^2-2|\nabla\phi|^2+(\Delta\phi)^2\right)}d\mathbf{\Omega},
\end{align*}
given that $\Psi_{c}$ and $\Psi_{e}$ are convex. From equation~\eqref{eq:sf2222}, we have that 
\begin{align}\label{eq:new}
\Psi_{c}'\left(\phi_{n+1}\right)-\Psi_{e}'\left(\phi_{n}\right)&=\tilde{\sigma}-\left(1+\Delta\right)\tilde{\theta}\nonumber \\
&+\dfrac{1}{2}\jump{\phi_n}\Psi_{c}''\left(\phi_{n+1}\right)+\dfrac{1}{2}\jump{\phi_n}\Psi_{e}''\left(\phi_{n}\right)+\alpha_n\Delta t \Delta\jump{\phi_n}.
\end{align}
We now simplify the notation for the explicit-implicit treatment of the second derivative and write
 \begin{align}\label{eq:lisandro3}
 \Psi''_{n,n+1}&=\Psi_{c}''\left(\phi_{n+1}\right)+\Psi_{e}''\left(\phi_{n}\right)\nonumber\\
 &=3\phi^2_{n+1} + \epsilon.
 \end{align}
We then multiply equation~\eqref{eq:new} by $\jump{\phi_n}=\phi_{n+1}-\phi_n$, use equation~\eqref{eq:sf2223}, and integrate over the domain, to obtain
\begin{align}\label{proof_2221}
\int_{\Omega}\left(\Psi_{c}'\left(\phi_{n+1}\right)-\Psi_{e}'\left(\phi_{n}\right)\right)\jump{\phi_n}d\mathbf{\Omega}&=\int_{\Omega}\left(\tilde{\sigma}\jump{\phi_n}-\left(1+\Delta\right)^2\{\phi_n\}\jump{\phi_n}\right)d\mathbf{\Omega}\nonumber\\
&+\int_{\Omega}\left(\dfrac{1}{2}\jump{\phi_n}\left(\Psi''_{n,n+1}\right)\jump{\phi_n}\right)d\mathbf{\Omega}\nonumber\\
&+\int_{\Omega}\left(\alpha_n\Delta t \Delta\jump{\phi_n}\jump{\phi_n}\right)d\mathbf{\Omega}.
\end{align}
We now proceed to expand the different terms on the right-hand side of equation~\eqref{proof_2221}.
Integrating the first term by parts, we obtain
\begin{align}
\int_{\Omega}\jump{\phi_n}\tilde{\sigma}d\mathbf{\Omega}&=\int_{\Omega}\Delta t\left(\Delta \tilde{\sigma}\right)\tilde{\sigma}d\mathbf{\Omega}\nonumber\\
&=-\int_{\Omega}\Delta t |\nabla \tilde{\sigma}|^2d\mathbf{\Omega}.\label{eq:proof21}
\end{align}
Then, for the second term in equation~\eqref{proof_2221}, we have
\begin{align}
\int_{\Omega}\left(1+\Delta\right)^2\{\phi_n\}\jump{\phi_n}d\mathbf{\Omega}&=\int_{\Omega}\left(1+\Delta\right)\{\phi_n\}\left(1+\Delta\right)\jump{\phi_n}d\mathbf{\Omega}\nonumber\\
&=\dfrac{1}{2}\int_{\Omega} \jump{\left(\phi_n\right)^2-2|\nabla\phi_n|^2+(\Delta\phi_n)^2}d\mathbf{\Omega}. \label{eq:proof22}
\end{align}
%
Using $\jump{\phi_{n}}=\Delta t \Delta \tilde{\sigma}$, taking the supremum, and integrating by parts the third term of equation~\eqref{proof_2221}, we have
\begin{align}
\int_{\Omega}\dfrac{1}{2}\jump{\phi_n}^2\Big(\Psi''_{n,n+1}\Big)d\mathbf{\Omega}\leq& \ \text{sup}\Big(\Psi''_{n,n+1}\Big) \int_{\Omega}\dfrac{\Delta t}{2}\jump{\phi_n}^2d\mathbf{\Omega} \nonumber\\
=& \ \text{sup}\Big(\Psi''_{n,n+1}\Big) \int_{\Omega}\dfrac{\Delta t}{2}\jump{\phi_n}\Delta \sigma d\mathbf{\Omega} \nonumber\\
=&-\dfrac{\Delta t}{2}\text{sup}\Big(\Psi''_{n,n+1}\Big)\int_{\Omega}\nabla \jump{\phi_n}\nabla\sigma d\mathbf{\Omega}. \label{eq:proof23}
\end{align}
\begin{remark}
Recalling equation~\eqref{eq:lisandro3}, $\Psi''_{n,n+1}=3\phi^2_{n+1} + \epsilon$ and the term is always positive. Thus, we may pull out a supremum without an absolute value needed. Moreover, since we assume the existence of a solution at each time step in $H^3(\Omega)$, such a supremum exists.
\end{remark}
\noindent Integrating the last term of equation~\eqref{proof_2221} by parts results in
\begin{align}
\int_{\Omega}\alpha_n\Delta t \Delta\jump{\phi_n}\jump{\phi_n}d\mathbf{\Omega}=-\int_{\Omega}\alpha_n\Delta t \left|\nabla\jump{\phi_n}\right|^2d\mathbf{\Omega}.\label{eq:proof24}
\end{align}
Finally, by collecting the terms in equations~\eqref{eq:proof21}-\eqref{eq:proof24}, and replacing them in equation~\eqref{proof_2221}, we obtain
\begin{align}
\jump{\mathcal{F}}\leq \int_{\Omega}\Big(&-\Delta t |\nabla \tilde{\sigma}|^2 - \alpha_n\Delta t\left(\nabla\jump{\phi_n}\right)^2\nonumber\\
&-\dfrac{\Delta t }{2}\text{sup}\Big(\Psi''_{n,n+1}\Big)\nabla\jump{\phi_n}\nabla \tilde{\sigma}\Big)d\mathbf{\Omega}.\label{eq:FE222proof}
\end{align}
Using Young's inequality, $2fg\leq\beta f^2+\beta^{-1}g^2$, with $f=-\nabla\jump{\phi_n}$ and $g=\nabla\tilde{\sigma}$, we then have that
\begin{align*}
\dfrac{\Delta t }{2}\text{sup}\Big(\Psi''_{n,n+1}\Big)(-\nabla\jump{\phi_n})\nabla \tilde{\sigma}
\leq&\dfrac{\Delta t }{4}\text{sup}\Big(\Psi_{n,n+1}''\left(\phi_{}\right)\Big)\left(\beta\left(\nabla\jump{\phi_n}\right)^2+\dfrac{|\nabla\tilde{\sigma}|^2}{\beta}\right),
\end{align*}
such that inequality~\eqref{eq:FE222proof} becomes
\begin{align}\label{eq:lisandro1}
\jump{\mathcal{F}}\leq\int_{\Omega}\Bigg(&-\Delta t |\nabla \tilde{\sigma}|^2-\alpha_n\Delta t \left(\nabla\jump{\phi_n}\right)^2+\dfrac{\Delta t }{4\beta}\text{sup}\left(\Psi_{n,n+1}''\left(\phi_{}\right)\right)|\nabla \tilde{\sigma}|^2 \nonumber\\
&+ \dfrac{\Delta t }{4}\text{sup}\left(\Psi_{n,n+1}''\left(\phi_{}\right)\right)\beta(\nabla\jump{\phi_n})^2\Bigg)d\mathbf{\Omega},
\end{align}
which is verified as long as
\begin{align}\label{eq:lisandro2}
\beta\geq\dfrac{\text{sup}\left(\Psi''_{n,n+1}\right)}{4}\quad \text{and} \quad \alpha_n\geq\dfrac{\left(\text{sup}\left(\Psi_{n,n+1}''\right)\right)^2}{16}.
\end{align}
Equation~\eqref{eq:lisandro1} and the fulfilment of the conditions in~\eqref{eq:lisandro2}, guarantee free energy stability.

\begin{remark}
The above condition is effectively nonlinear, since the choice of $\alpha_n$ depends on $\phi_{n+1}$. However, since the smoothness at each time step is assumed to be $\mathcal{H}^3$, it is continuous and a global supremum of $\Psi''_{n,n+1}\!=\!3\phi^2_{n+1} + \epsilon$ exists at each time step. The supremum of such a quantity is however a priori unknown. Thus, in our implementation  the above stability condition is a lagging condition where $\alpha_n$ is computed using the current time step.  Another approach involves truncating the second derivative of $\Psi$ outside the regions $[-1,1]$, and interpolating with polynomials as in \cite{kris} in the context of Cahn--Hilliard to obtain a global bound, such that $\alpha_n$ can be evaluated independently from $\Delta t$. 
\end{remark}

\subsubsubsection{Alternative formulation}
This stabilization procedure is also suitable for the following alternative formulation
\begin{subequations}
\begin{align}
\frac{\partial \phi}{\partial t}&=\Delta \sigma&  \text{\space \space  in } \Omega\times]0,T],\label{HG1}\\
\sigma&=\left(1+\Delta\right)^2\phi + \Psi'(\phi) &\text{\space \space  in } \Omega\times]0,T].\label{HG2}
\end{align}
\end{subequations}
Let us denote by $\mathcal{V}_2$ a functional space belonging to~$\mathcal{H}^2$, where~$\mathcal{H}^2$ is the Sobolev space of square-integrable functions with square-integrable first and second derivatives. Assuming periodic boundary conditions in all directions, a weak form can be derived multiplying~\eqref{HG1}-\eqref{HG2} by test functions $q,\!~w\!~\!\in\!~\!~\mathcal{V}_2$, respectively, and integrating the equations by parts. The problem can then be defined as that of finding~$\phi,\!~\!\sigma\!~\!\in\!~\!\mathcal{V}_2$ such that for all~$q, w \in \mathcal{V}_2$
\begin{align}
0&=\left(q,\dot{\phi}\right)_{\Omega} + (\nabla q,\nabla \sigma)_{\Omega}\nonumber\\
&+\left(w,\sigma-\Psi'\left(\phi\right)-\phi\right)_{\Omega}+2\left(\nabla w,\nabla \phi\right)_{\Omega}-\left(\Delta w, \Delta \phi\right)_{\Omega}.
\end{align}
This formulation requires the use of at least~$\!\mathcal{C}^1\!$~continuity, but the use of a convex-implicit and concave-explicit discretization of the nonlinearity can also be done, such that the fully discrete formulation becomes
\begin{align*}
0&=\left(q^h,\frac{\jump{ \phi_{n}^h}}{\Delta t}\right)_{\Omega} + \left(\nabla q^h, \nabla \sigma^h\right)_{\Omega}+\left(w^h,\sigma^h-\{\phi_n^h\}\right)_{\Omega}\nonumber\\
&-\left(w^h,\Psi'_{c}\left(\phi^h_{n+1}\right)-\Psi_{c}''\left(\phi^h_{n+1}\right)\frac{\jump{\phi_n^h}}{2}\right)_{\Omega}\nonumber\\
&+\left(w^h,\Psi'_{e}\left(\phi^h_{n}\right)+\Psi_{e}''\left(\phi^h_{n}\right)\frac{\jump{\phi_n^h}}{2}\right)_{\Omega}\nonumber\\
&+\left(\nabla w^h,2\nabla \{\phi_n^h\} -\alpha_n\Delta t \nabla \jump{\phi_n^h}\right)_{\Omega} - \left(\Delta w^h, \Delta \{\phi_n^h\}\right)_{\Omega},
\end{align*}
where 
\begin{itemize}
\item $\alpha_n\geq\dfrac{\left[\text{sup}\left(\Psi''_{n,n+1}\right)\right]^2}{16}$,\\ 
\item $\Psi''_{n,n+1}=3\phi^2_{n+1}+\epsilon$.
\end{itemize}
\subsubsection{Numerical implementation}
With regards to the implementation, we let the global vectors of degrees of freedom associated to $\phi^h_n$, $\sigma^h_n$ and $\theta^h_n$ be $\mathbf{\Phi}_n$, $\mathbf{\Sigma}_n$ and $\mathbf{\Theta}_n$, respectively. The residual vectors for this formulation are then given by
\begin{align*}
\mathbf{R}_{\phi}(\mathbf{\Phi}_n,\mathbf{\Phi}_{n+1},\mathbf{\Sigma}_{n+1},\mathbf{\Theta}_{n+1});\quad \mathbf{R}_{\phi}=\{R_A^\phi\};\quad A=1,...,n_b,\\
\mathbf{R}_{\sigma}(\mathbf{\Phi}_n,\mathbf{\Phi}_{n+1},\mathbf{\Sigma}_{n+1},\mathbf{\Theta}_{n+1});\quad \mathbf{R}_{\sigma}=\{R_A^\sigma\};\quad A=1,...,n_b, \\
\mathbf{R}^{\theta}(\mathbf{\Phi}_n,\mathbf{\Phi}_{n+1},\mathbf{\Sigma}_{n+1},\mathbf{\Theta}_{n+1});\quad \mathbf{R}_{\theta}=\{R_A^\theta\};\quad A=1,...,n_b,
\end{align*}
where
\begin{align*}
R_A^\phi&= \left(N_A,\frac{\jump{ \phi_{n}^h}}{\Delta t}\right) + \left(\nabla N_A, \nabla \sigma^h\right),\\
R_A^\sigma&=\left(N_A,\sigma^h-\theta^h-\left(\Psi_{c}'\left(\phi^h_{n+1}\right)-\Psi_{c}''\left(\phi^h_{n+1}\right)\frac{\jump{\phi_n^h}}{2}\right)\right)\nonumber\\
&+\left(N_A,\left(\Psi_{e}'\left(\phi^h_{n}\right)+\Psi_{e}''\left(\phi^h_{n}\right)\frac{\jump{\phi_n^h}}{2}\right)\right)\nonumber\\
&+\left(\nabla N_A,\nabla \theta^h -\alpha_n\Delta t \nabla \jump{\phi_n^h}\right),\\
R_A^\theta&=\left(N_A,\theta^h-\{\phi_n^h\} \right) + \left(\nabla N_A,\nabla \{\phi_n^h\} \right).
\end{align*}
The resulting system of nonlinear equations for $\mathbf{\Phi}_{n+1}$, $\mathbf{\Sigma}_{n+1}$ and $\mathbf{\Theta}_{n+1}$ is solved using Newton's method, where~$\mathbf{\Phi}_{n+1}^{(i)}$,~$\mathbf{\Sigma}_{n+1}^{(i)}$ and $\mathbf{\Theta}_{n+1}^{(i)}$ correspond to the $\it{i}$-$\it{th}$ iteration of Newton's algorithm. The iterative procedure is specified in Algorithm~\ref{algorithm3}.
\begin{algorithm}
Taking $\mathbf{\Phi}_{n+1}^{(0)}=\mathbf{\Phi}_{n}$, $\mathbf{\Sigma}_{n+1}^{(0)}=\mathbf{\Sigma}_{n}$, and $\mathbf{\Theta}_{n+1}^{(0)}=\mathbf{\Theta}_{n}$, for $i=1,..,\text{i}_{max}$,
\\\\
(1) Compute the residuals $\mathbf{R}_{\phi}^{(i)}$, $\mathbf{R}_{\sigma}^{(i)}$, $\mathbf{R}_{\theta}^{(i)}$, using $\mathbf{\Phi}_{n+1}^{(i)}$, $\mathbf{\Sigma}_{n+1}^{(i)}$, $\mathbf{\Theta}_{n+1}^{(i)}$.\\
(2) Compute the Jacobian matrix $\mathbf{K}^{(i)}$ using the $\it{i}$-$\it{th}$ iterates. This matrix is given by
\begin{align}\label{jacobian_components222}
\mathbf{K}^{(i)}= \begin{pmatrix}
\mathbf{K}^{\phi\phi}        & \mathbf{K}^{\phi\sigma}      & \mathbf{K}^{\phi\theta}\\
\mathbf{K}^{\sigma\phi}   & \mathbf{K}^{\sigma\sigma}  & \mathbf{K}^{\sigma\theta}\\
\mathbf{K}^{\theta\phi}   & \mathbf{K}^{\theta\sigma}  & \mathbf{K}^{\theta\theta}
     \end{pmatrix}^{(i)},
\end{align}
where the individual components of each submatrix of the Jacobian are defined in the Appendix in equations~(\ref{jac22200}) through~(\ref{jac22222}).

(3) Solve the linear system 
\begin{align}
\begin{pmatrix}
\mathbf{K}^{\phi\phi}        & \mathbf{K}^{\phi\sigma}      & \mathbf{K}^{\phi\theta}\\
\mathbf{K}^{\sigma\phi}   & \mathbf{K}^{\sigma\sigma}  & \mathbf{K}^{\sigma\theta}\\
\mathbf{K}^{\theta\phi}   & \mathbf{K}^{\theta\sigma}  & \mathbf{K}^{\theta\theta}
     \end{pmatrix}^{(i)}
     \begin{pmatrix}
\Delta\mathbf{\Phi}              \\
       \Delta\mathbf{\Sigma}  \\
       \Delta\mathbf{\Theta}
     \end{pmatrix}^{(i+1)}
     =
        \begin{pmatrix}
\mathbf{R}_{\phi}               \\
\mathbf{R}_{\sigma} \\
\mathbf{R}_{\theta} 
     \end{pmatrix}^{(i)}\nonumber.
     \end{align}
     (4) Update the solution such that
     \begin{align}
     \begin{pmatrix}
\mathbf{\Phi}_{n+1}               \\
       \mathbf{\Sigma}_{n+1}  \\
       \mathbf{\Theta}_{n+1}
     \end{pmatrix}^{(i+1)}
     =
     \begin{pmatrix}
\mathbf{\Phi}_{n+1}               \\
       \mathbf{\Sigma}_{n+1} \\ 
       \mathbf{\Theta}_{n+1}
     \end{pmatrix}^{(i)}
     -
     \begin{pmatrix}
\Delta\mathbf{\Phi}               \\
       \Delta\mathbf{\Sigma} \\
       \Delta\mathbf{\Theta} 
     \end{pmatrix}^{(i+1)}\nonumber.
     \end{align}
     \caption{Iterative procedure to solve the $2+2+2$ mixed form}
     Steps (1) through (4) are repeated until the norms of the global residual vector are reduced to a certain tolerance ($10^{-8}$ in all the examples shown in this work) of their initial value. Convergence is usually achieved in 2 or 3 nonlinear iterations per time step.
\label{algorithm3}
\end{algorithm}

\section{Numerical results}\label{s:cost}
The implementation of the numerical scheme described in section~\ref{s:numfor} was done using PetIGA~\cite{petiga,framework,Cortes}, which is a software framework built on top of PETSc~\cite{petsc1,petsc2}, that delivers a high-performance computational framework for IGA. Tutorials for the framework are being developed and can be found in~\cite{Collier}. 
This section describes the calculation of the free energy for the discretization, presents numerical evidence to verify the results in section~\ref{subset:time} in two dimensions, and shows the performance of the method on some more challenging three-dimensional problems related to the growth of crystals in a supercooled liquid.
\subsection{Free-energy computation}
If one uses spaces that are at least $\mathcal{C}^1$-continuous, the free energy can be computed as
\begin{align*}
\mathcal{F}[\phi_n^h]=\int_{\Omega}\left[\Psi(\phi_n^h)+\frac{1}{2}\left(\left(\phi_n^h\right)^2-2|\nabla\phi_n^h|^2+(\Delta\phi_n^h)^2\right) \right]d\mathbf{\Omega}.
\end{align*}
Modifications are needed in the $2+2+2$ case though, as the discrete atomistic density $\phi^h$ only lives in $\mathcal{H}^1$. As such, $\Delta \phi^h_n$ is undefined. This obstacle can be overcome by making use of the auxiliary variable $\theta$, as
\begin{align*}
\theta = \left(1+ \Delta\right)\phi \quad \Leftrightarrow \quad \Delta \phi = \theta-\phi,
\end{align*}
such that the free energy functional can be computed as
\begin{align*}
\mathcal{F}[\phi_n^h]=\int_{\Omega}\left[\Psi(\phi_n^h)+\frac{1}{2}\left(\left(\phi_n^h\right)^2-2|\nabla\phi_n^h|^2+(\theta_n^h-\phi_n^h)^2\right) \right]d\mathbf{\Omega}.
\end{align*}
\begin{remark}
The use of the auxiliary variables means that they also have to be initialised, as the initial condition is only specified for $\phi$. A nonlinear ${L}_2$ projection is performed to solve the semidiscrete versions of equations~\eqref{triple2} and \eqref{triple3}, shown in the last two lines of equation~\eqref{222}. 
\end{remark}
\subsection{Numerical validation of the stable scheme}\label{sec:validation}
\begin{figure}
\centering
{\includegraphics[width=0.5\textwidth]{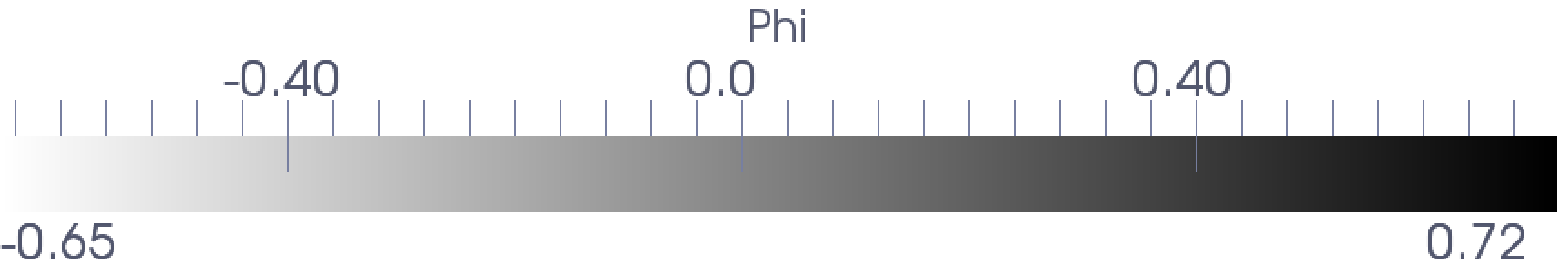}}
\begin{tabular}{ccc}
{\includegraphics[width=0.3\textwidth]{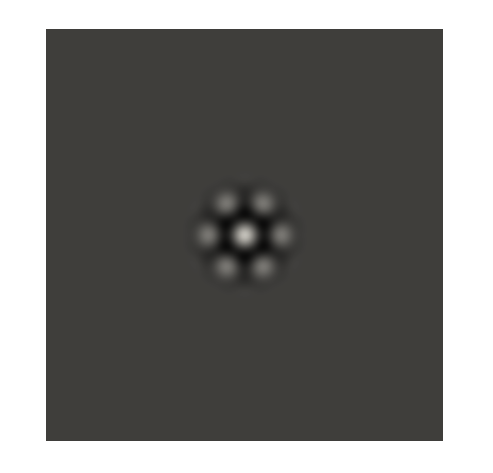}\label{fig:0}}&
{\includegraphics[width=0.3\textwidth]{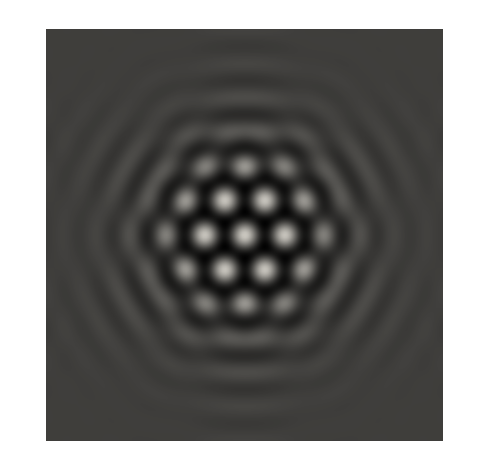}\label{fig:1}} &
{\includegraphics[width=0.3\textwidth]{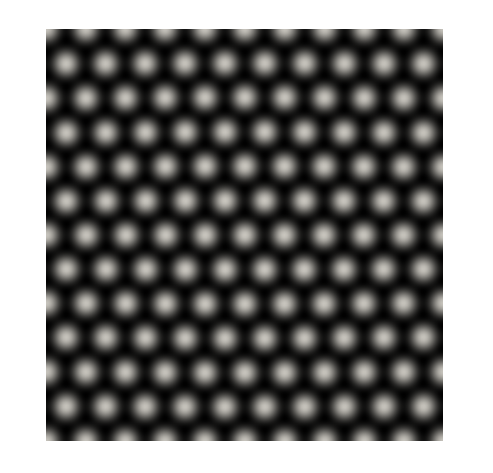}\label{fig:2}} \\
$t=0\quad$&$t=25\quad$&$t=150$
\end{tabular}
\caption{Snapshots of the approximate dimensionless atomistic density field showing its evolution throughout the simulation, which was run using a computational mesh composed of $256\times266$~$\mathcal{C}^0$ linear elements, with a time step size of $1.0$.}
\label{fig:snapshots1}
\end{figure}

\begin{figure}
\begin{center}
\begin{tikzpicture}
  \node (img1) {\includegraphics[width=1.0\textwidth]{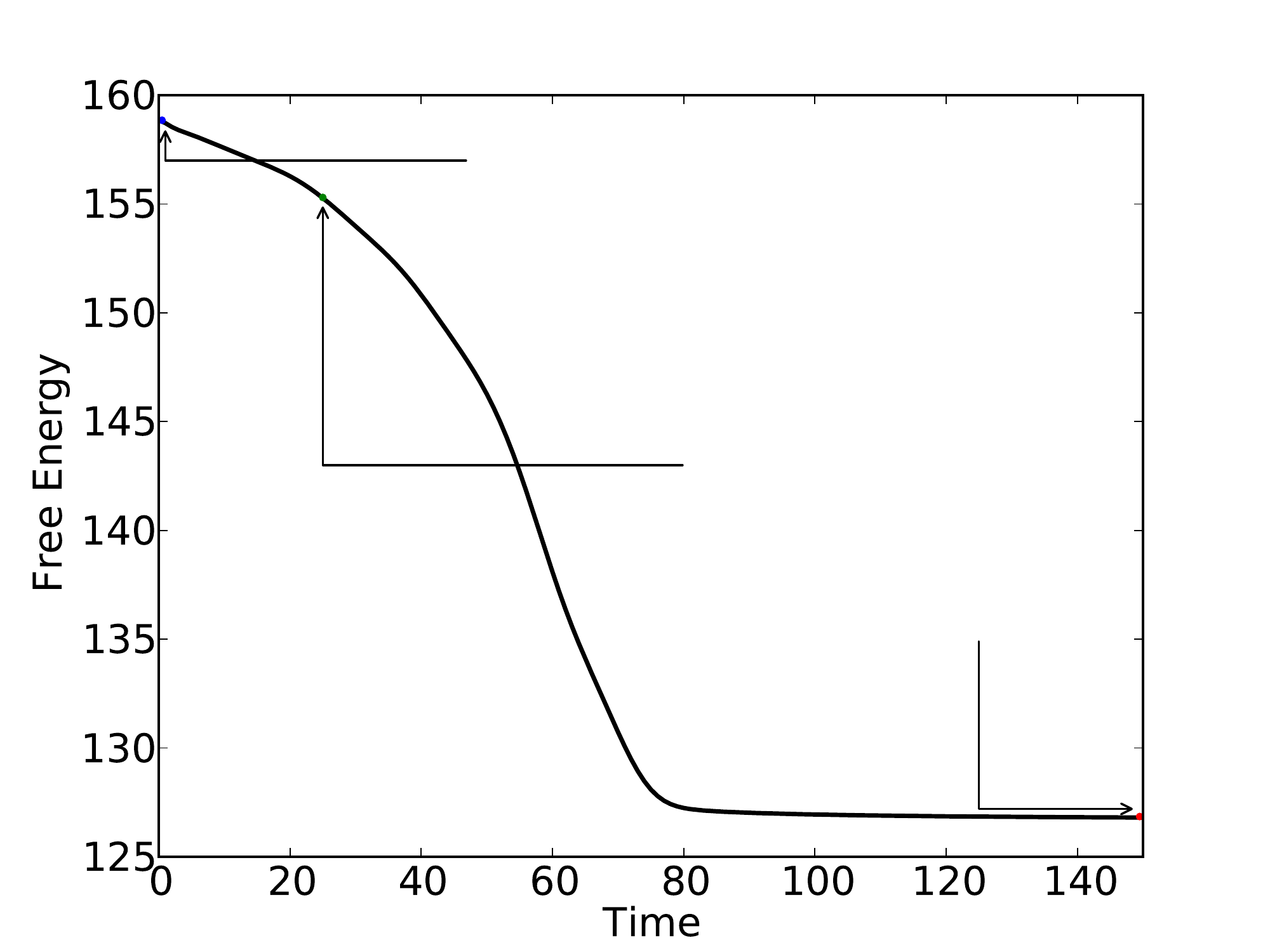}};
  \node (img2) at (img1.center)[yshift = 2.7cm][xshift=-0.5cm] {\includegraphics[width=0.2\textwidth]{fig-pfc0_bis.png}};
  \node (img3) at (img1.center)[yshift = 0cm][xshift=1cm] {\includegraphics[width=0.2\textwidth]{fig-pfc25_bis.png}};
  \node (img3) at (img1.center)[yshift = -2cm][xshift=4cm] {\includegraphics[width=0.2\textwidth]{fig-pfc150_bis.png}};
\end{tikzpicture}
\caption{Free energy evolution. The free energy is monotonically decreasing throughout the simulation, which was run using a computational mesh composed of $256\times266$~$\mathcal{C}^0$ linear elements. A time step size of $1$ was used, with an $\alpha_n$ value of~0.25.}\label{fig:fe_evolution}
\end{center}
\end{figure}

\begin{figure}
\centering
\includegraphics[width = 1.0\textwidth]{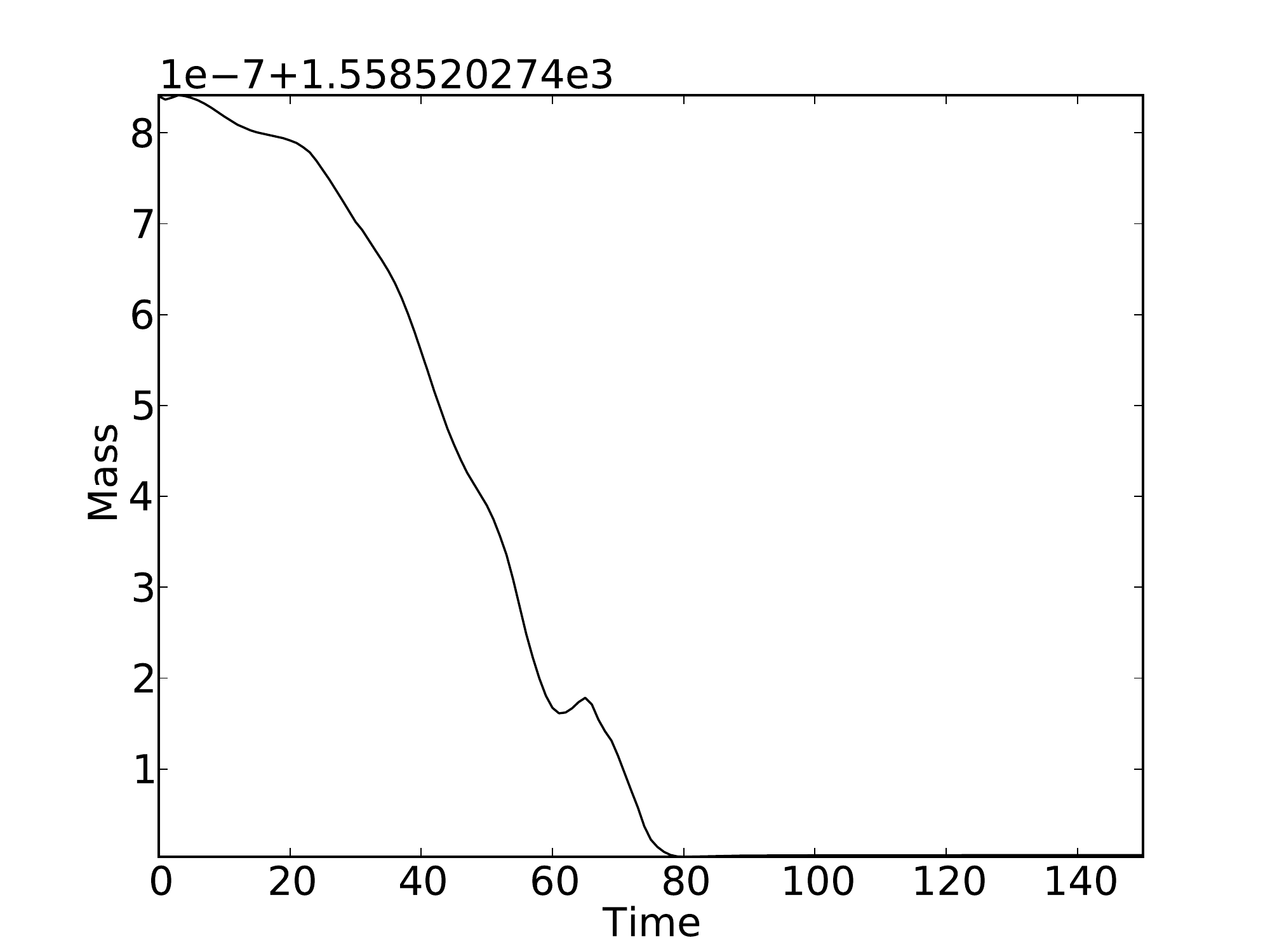}
\caption{Mass evolution. The changes in mass are below the criterion for numerical convergence, which validates numerically that mass is indeed conserved. The error can be attributed to quadrature as well as the iterative solver.}\label{fig:m_evolution} 
\end{figure}

As a test example, we simulate the two-dimensional growth of a crystal in a supercooled liquid, using one-mode approximations for the density profiles of the crystalline structures~\cite{Elder0,Elder1}. The one-mode approximation corresponding to a triangular configuration is defined as
\begin{align}\label{eq:phi_s}
\phi_S\left(\mathbf{x}\right)=\text{cos}\left(\dfrac{q}{\sqrt{3}}y\right)\text{cos}\left(qx\right)-\dfrac{1}{2}\text{cos}\left(\dfrac{2q}{\sqrt{3}}y\right),
\end{align}
where $q$ represents a wavelength related to the lattice constant~\cite{Elder2010}, and $x$ and $y$ represent the Cartesian coordinates. A solid crystallite is initially placed in the centre of a liquid domain, which is assigned an average density $\bar{\phi}$. The initial condition becomes
\begin{align}\label{eq:initial}
\phi_0\left(\mathbf{x}\right)=\bar{\phi}+\omega(\mathbf{x})\left(A\phi_S\left(\mathbf{x}\right)\right),
\end{align}
where $A$ represents an amplitude of the fluctuations in density, and the scaling function $\omega(\mathbf{x})$ is defined as
\begin{equation*}
\omega(\mathbf{x})=
\begin{cases}
\left(1-\left(\dfrac{||\mathbf{x}-\mathbf{x}_0||}{d_0}\right)^2\right)^2 &\text{if} \quad ||\mathbf{x}-\mathbf{x}_0|| \leq d_0\\
0 & \text{otherwise} 
\end{cases}
\end{equation*}
where $\mathbf{x}_0$ is the coordinate of the centre of the domain, and $d_0$ is $^1/_6$ of the domain length in the $x$-direction. Different lattices can be reproduced, depending on the values used for $\epsilon$ and the average atomistic density $\bar{\phi}$. Phase diagrams have been developed in both two~\cite{Elder2010} and three~\cite{Jaatinen} dimensions. In order to avoid mismatches on the boundaries when the grain boundaries meet, the computational domain ${\Omega}$ is dimensioned in such a way as to make it periodic along both directions. To do this while keeping the problem within a reasonable size, we use the frequency present in equation~\eqref{eq:phi_s} to define the domain ${\Omega}$ as 
\begin{align*}
{\Omega}=\Bigg[0,\dfrac{2\pi}{q}a\Bigg]\times\Bigg[0,\dfrac{\sqrt{3}\pi}{q}b\Bigg], 
\end{align*}
where $a$ and $b$ are assigned values of $10$ and $12$, respectively. These numbers are chosen so that the domain is almost square. The number of elements in the $y$-direction, $N_y$, is then defined as
\begin{align*}
N_y=\left\lfloor\dfrac{b\sqrt{3}}{2a}N_x+\dfrac{1}{2}\right\rfloor,
\end{align*}
where $N_x$ represents the number of elements in the $x$-direction. This adjustment is made to account for the difference in length between both directions, and to have the element size $h$ in both directions be approximately equal. The variables $q$ and $A$ are assigned their corresponding equilibrium values, obtained by minimizing the free energy presented in equation~\eqref{FEright}, with respect to both $A$ and $q$, while using the approximation of equation~\eqref{eq:phi_s} to define the atomistic density. For the results presented in this section, the values used are
\begin{align*}
\epsilon=0.325, \quad \bar{\phi}=\dfrac{\sqrt{\epsilon}}{2}, \quad A=\dfrac{4}{5}\left(\bar{\phi}+\dfrac{\sqrt{15\epsilon-36\bar{\phi}^2}}{3}\right), \quad q=\dfrac{\sqrt{3}}{2}.
\end{align*}
The parameter $\epsilon$ is chosen such that the triangular structure is stable~\cite{Elder2010,Jaatinen}. Snapshots of the simulation are shown in Fig.~\ref{fig:snapshots1}. The initial crystallite placed in the centre of the domain grows at the expense of the supercooled liquid, a state which is enforced by the degree of undercooling~$\epsilon$.
The non-increasing free energy and mass conservation, properties that need to be verified for a numerical scheme to be valid when solving this equation, can be verified in Figs.~\ref{fig:fe_evolution} and~\ref{fig:m_evolution}, respectively. 

This same example was used to perform the numerical validation of the results presented in section~\ref{subset:time}. The stabilization term $\alpha_n$ was assigned a value of $0.25$ for the range of time and space resolutions covered in this section. This choice of $\alpha_n$ was made after \textit{a-priori} numerical experimentation for this specific example using equations~\eqref{eq:lisandro2} and~\eqref{eq:lisandro3}.

\begin{figure}
\begin{center}
\begin{tabular}{c}
{\includegraphics[width=0.75\textwidth]{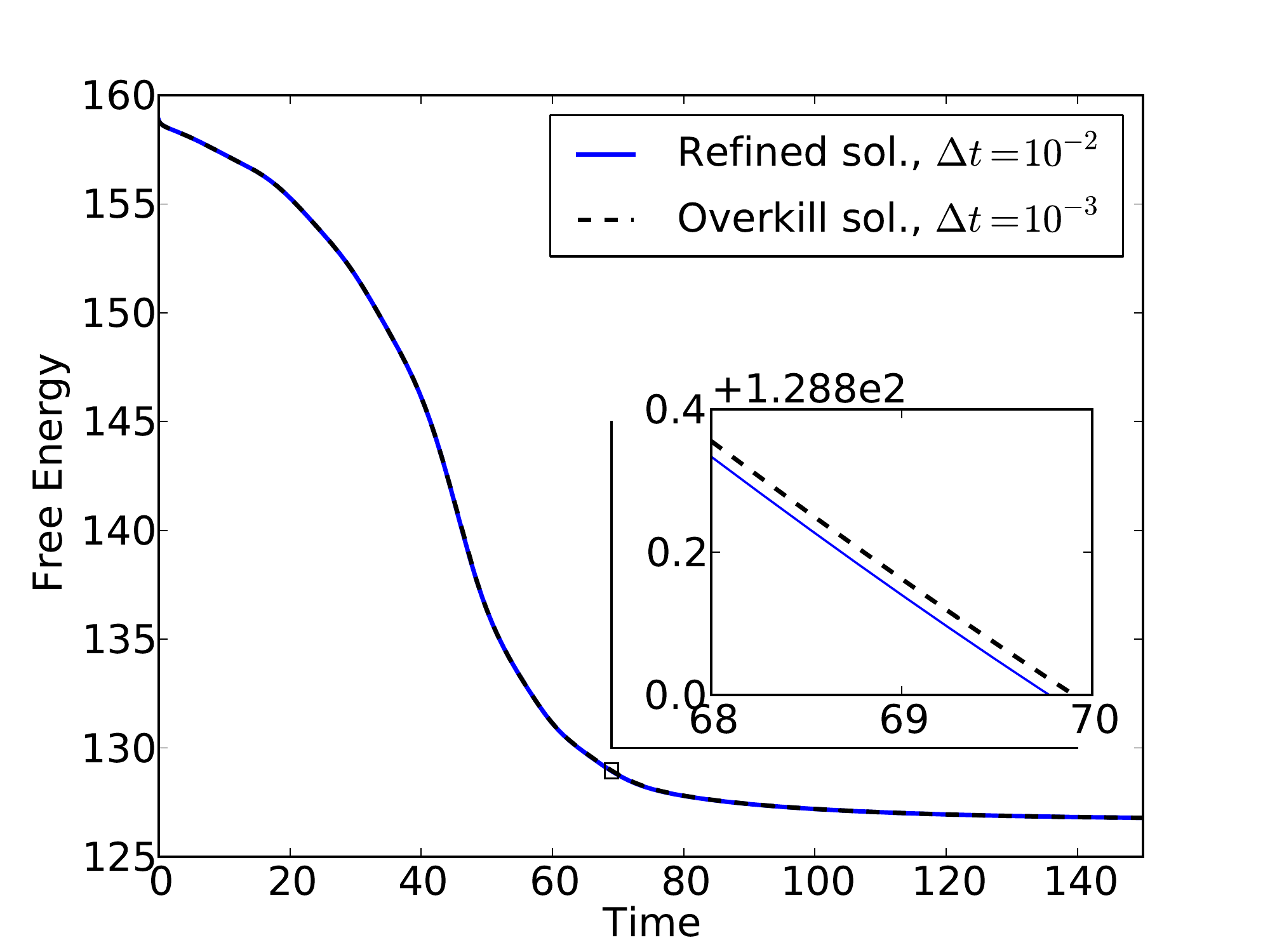}} \\
(a) Free energy evolution of overkill and reference solutions. \\
{\includegraphics[width=0.75\textwidth]{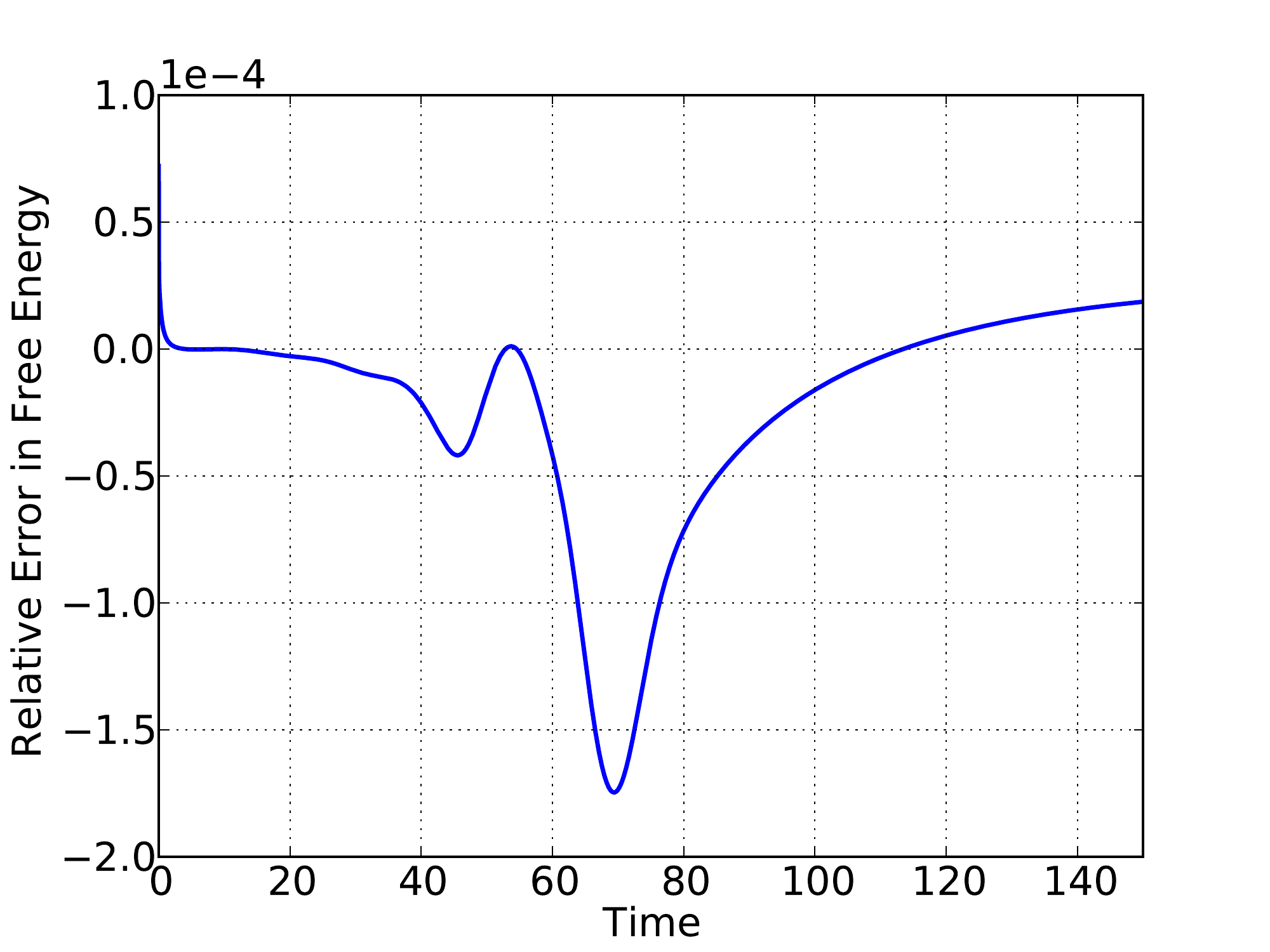}} \\
(b) Relative error between reference and overkill free energy evolutions. \\
\end{tabular}
\caption{Free energy evolutions of reference and overkill solutions. In (a), the free energy evolution of an overkill solution using $\left[128 \times133\right]$ quartic $\mathcal{C}^0$ elements and a time step size of $10^{-3}$ is shown along with the free energy evolution corresponding to a reference solution obtained using $\left[128 \times 133\right]$ quadratic $\mathcal{C}^0$ elements and a time step size of $10^{-2}$. An inset plot is shown on the bottom right corner of (a), in the region where the error is highest throughout the simulation as can be verified in (b), where the relative error between the reference and overkill free energy evolutions is shown.}
\label{fig:negligible}
\end{center}
\end{figure}

In order to study the convergence in time of the proposed method, a reference solution is required. We obtain this reference solution using a grid with $\left[128 \times 133\right]$ elements, $p^2\mathcal{C}^0$ basis functions, and a time step size $\Delta t=~10^{-2}$. This solution was obtained within a matter of hours using a workstation with $32$ processor cores. The order of the basis function $p$ was elevated in the case of this solution, as it is a more sensible choice than going for $h$-refinement with $p^1\mathcal{C}^0$ basis functions, as shown in~\cite{Vignal1}. Then, to assess the quality of this reference solution in terms of the error in the free energy, an overkill solution was calculated using a grid with $\left[128 \times 133\right]$ elements, $p^4\mathcal{C}^0$ spaces, and an order of magnitude smaller time step size $\Delta t=10^{-3}$. Using the same machine as before, the overkill solution took a week and a half to be completed. The free energy evolutions of the reference and overkill solutions are compared in Fig.~\ref{fig:negligible}a, while the relative error evolution between the free energies is shown in Fig.~\ref{fig:negligible}b. This comparison allows us to conclude that the reference solution is refined enough in space and time to proceed with the study of convergence in time.

\begin{figure}
\centering
{\includegraphics[width=0.9\textwidth]{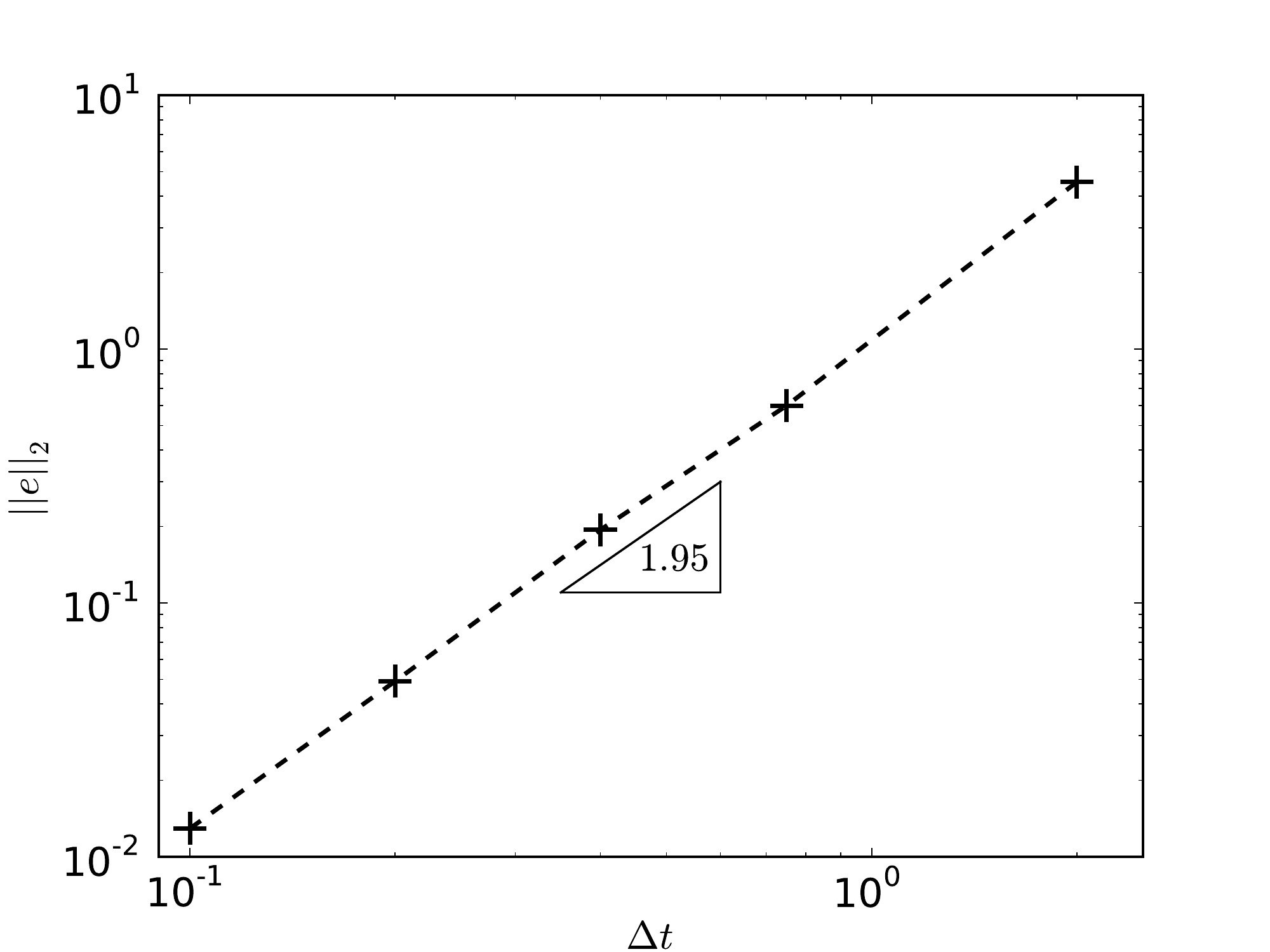}}
\caption{Log of $\mathcal{L}^2$-norm of the error at time $T=150$ versus the log of time step size $\Delta t$. The value of the slope confirms the method is second-order accurate in time. The mesh used was made up of $\left[128 \times 133\right]$ quadratic $\mathcal{C}^0$ elements, such that the spatial error could be considered negligible in the simulations. The parameter $\alpha_n$ was given a value equal to $0.25$, which complies with the bound presented in section~\ref{subset:time} for this problem.}
\label{fig:convergence}
\end{figure}

\begin{figure}
\centering
{\includegraphics[width=0.9\textwidth]{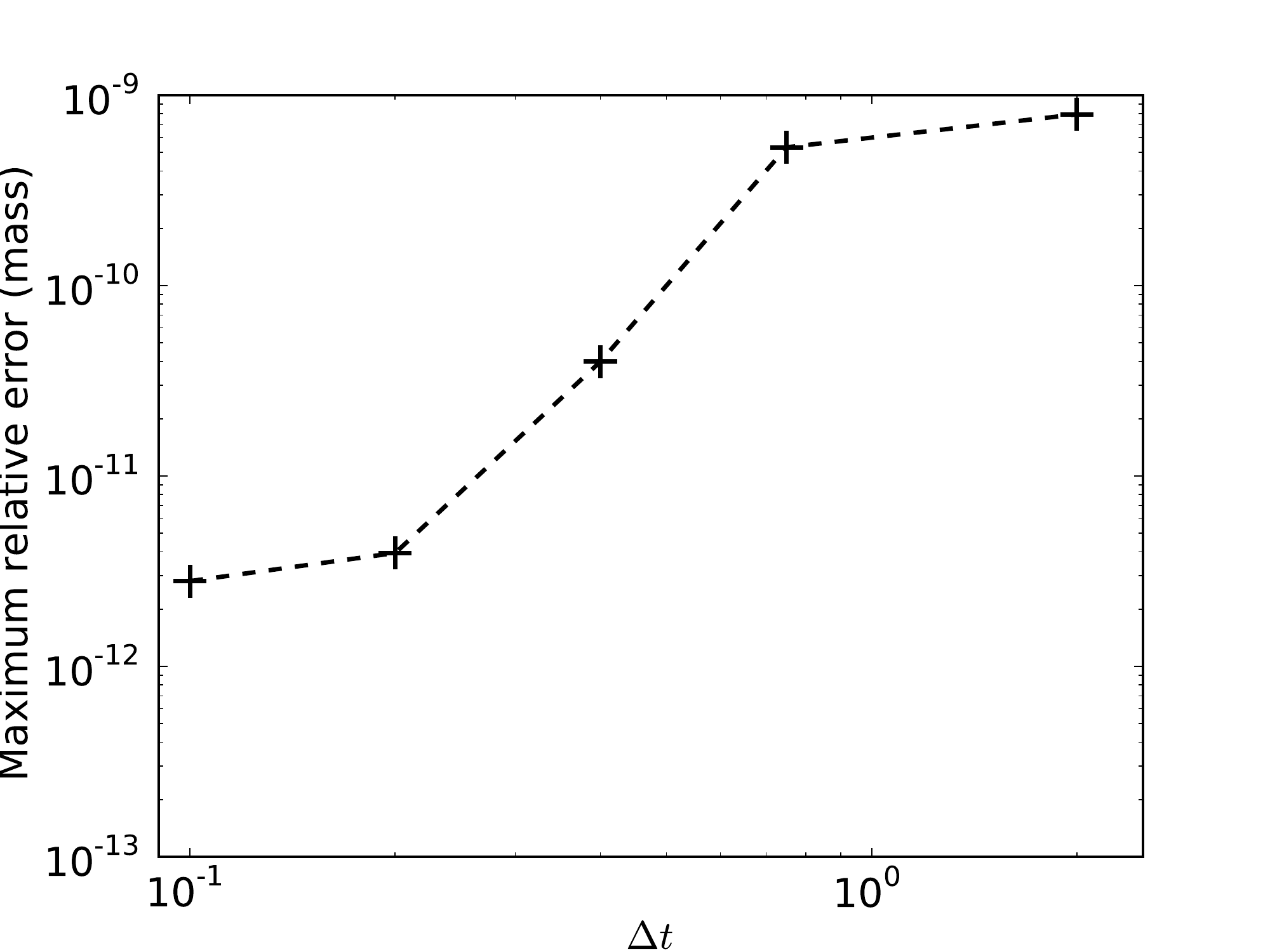}}
\caption{Mass conservation. The maximum relative error over the entire evolution of the system remained below $10^{-9}$ for the simulations considered in this work. The mesh used was made up of $\left[128 \times 133\right]$ quadratic $\mathcal{C}^0$ elements, such that the spatial error could be considered negligible in the simulations. The parameter $\alpha_n$ was given a value equal to $0.25$, which complies with the bound presented in section~\ref{subset:time} for this problem.}
\label{fig:conservation}
\end{figure}

\begin{figure}
\centering
{\includegraphics[width=0.9\textwidth]{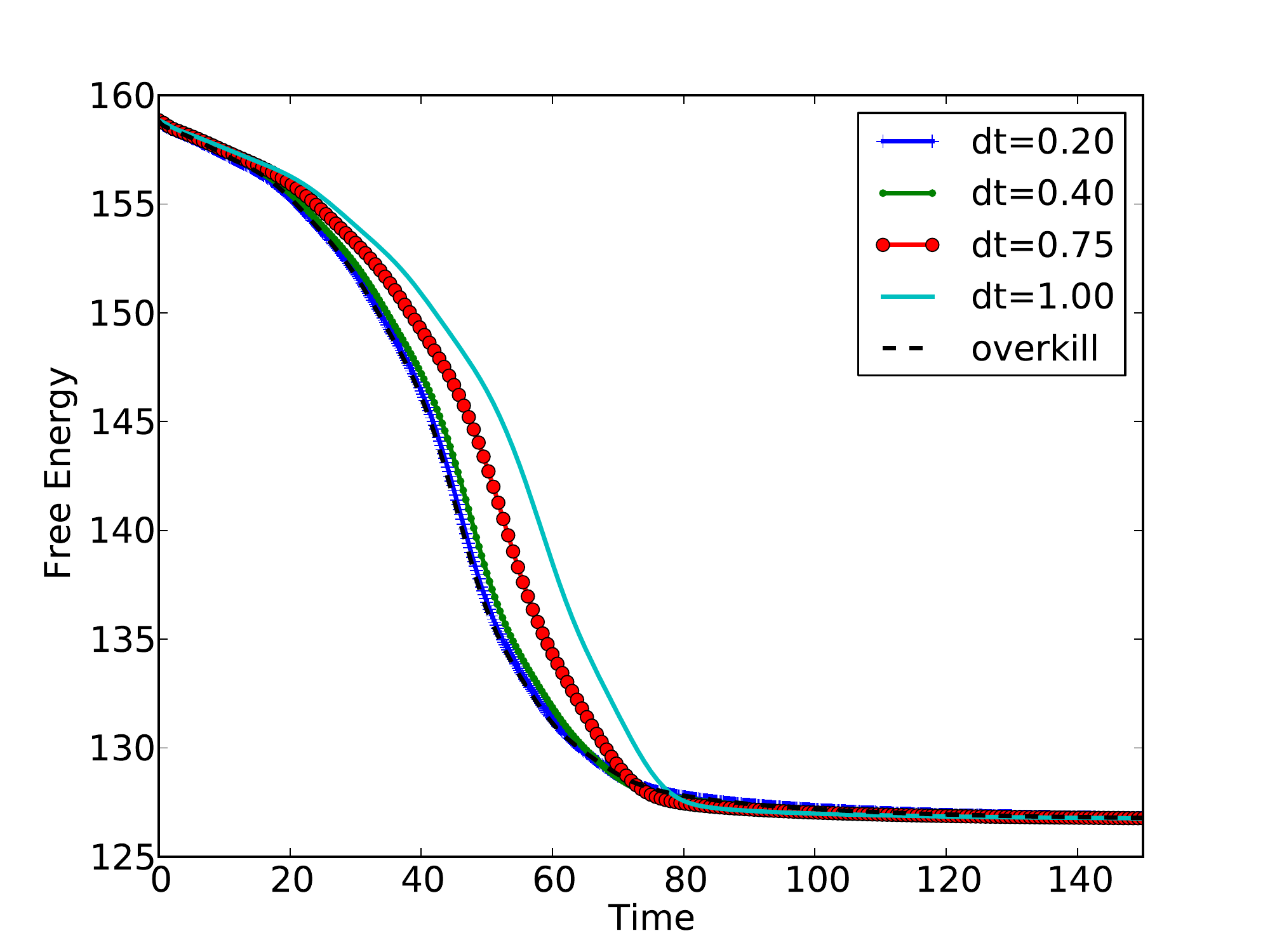}}
\caption{Free energy monotonicity. The free energy functional of the system exhibits strong energy stability, such that $\mathcal{F}\left[\phi\left(t_{n+1}\right)\right]\leq \mathcal{F}\left[\phi\left(t_{n}\right)\right]$. This is independent of the time step size used as can be observed in the plot. The mesh used was made up of $\left[128 \times 133\right]$ quadratic $\mathcal{C}^0$ elements, such that the special error could be considered negligible in the simulations. The parameter $\alpha_n$ was given a value equal to $0.25$, which complies with the bound presented in section~\ref{subset:time} for this problem.}
\label{fig:energy}
\end{figure}

\begin{figure}
\centering
\begin{tabular}{cc}
 {\includegraphics[trim=0cm 0.3cm 1.4cm 1.3cm, clip=true,width=0.48\columnwidth]{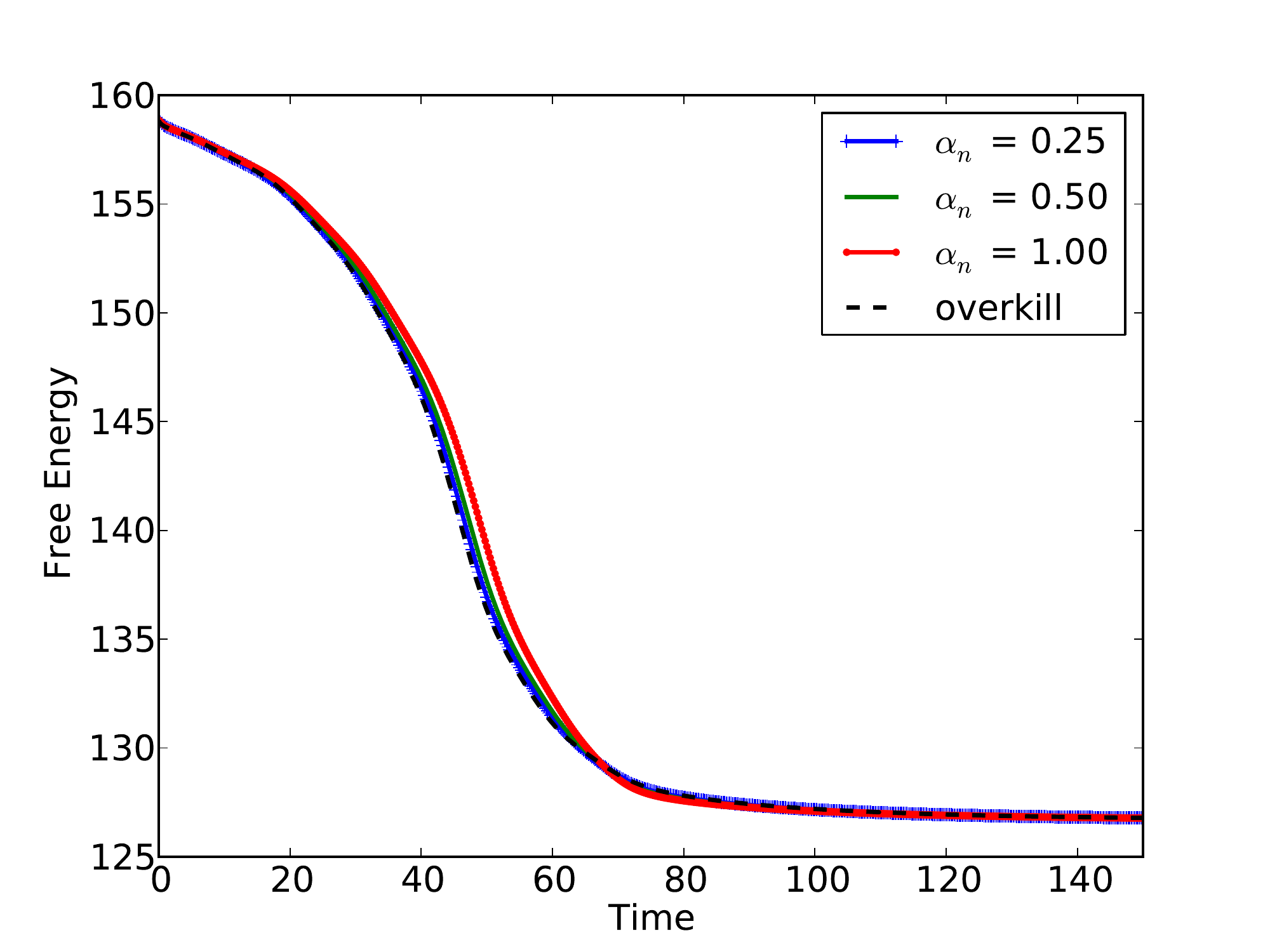}\label{pfc2d_dt0_25}}~\!~
 {\includegraphics[trim=0cm 0.3cm 1.4cm 1.3cm, clip=true,width=0.48\columnwidth]{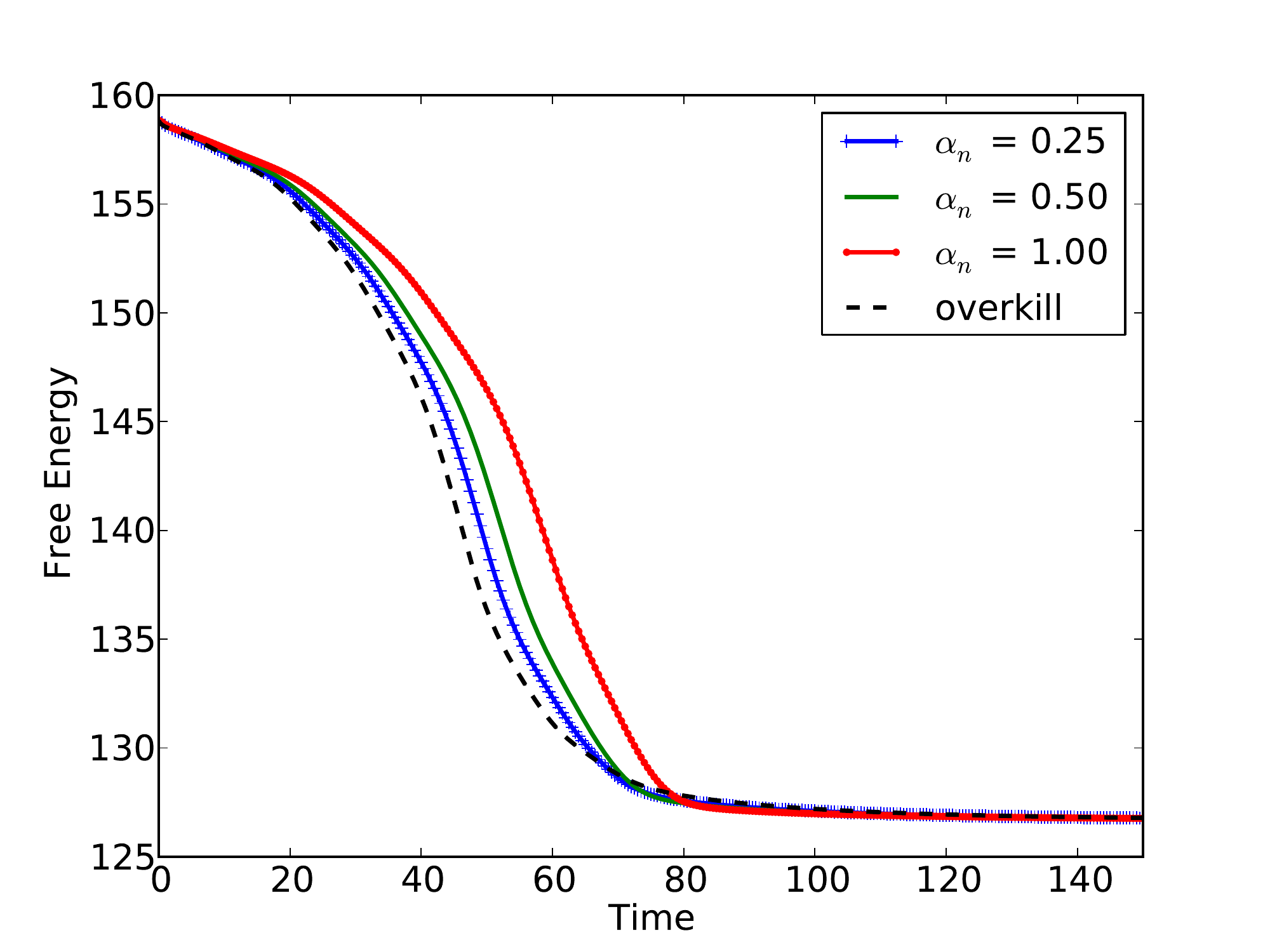}}\label{pfc2d_dt0_5} \\
\qquad \qquad(a) $\Delta t=0.25$ \qquad \qquad \qquad \qquad \qquad (b) $\Delta t=0.5$\qquad \vspace{0.5cm}\\
{\includegraphics[trim=0cm 0.3cm 1.4cm 1.3cm, clip=true,width=0.48\columnwidth]{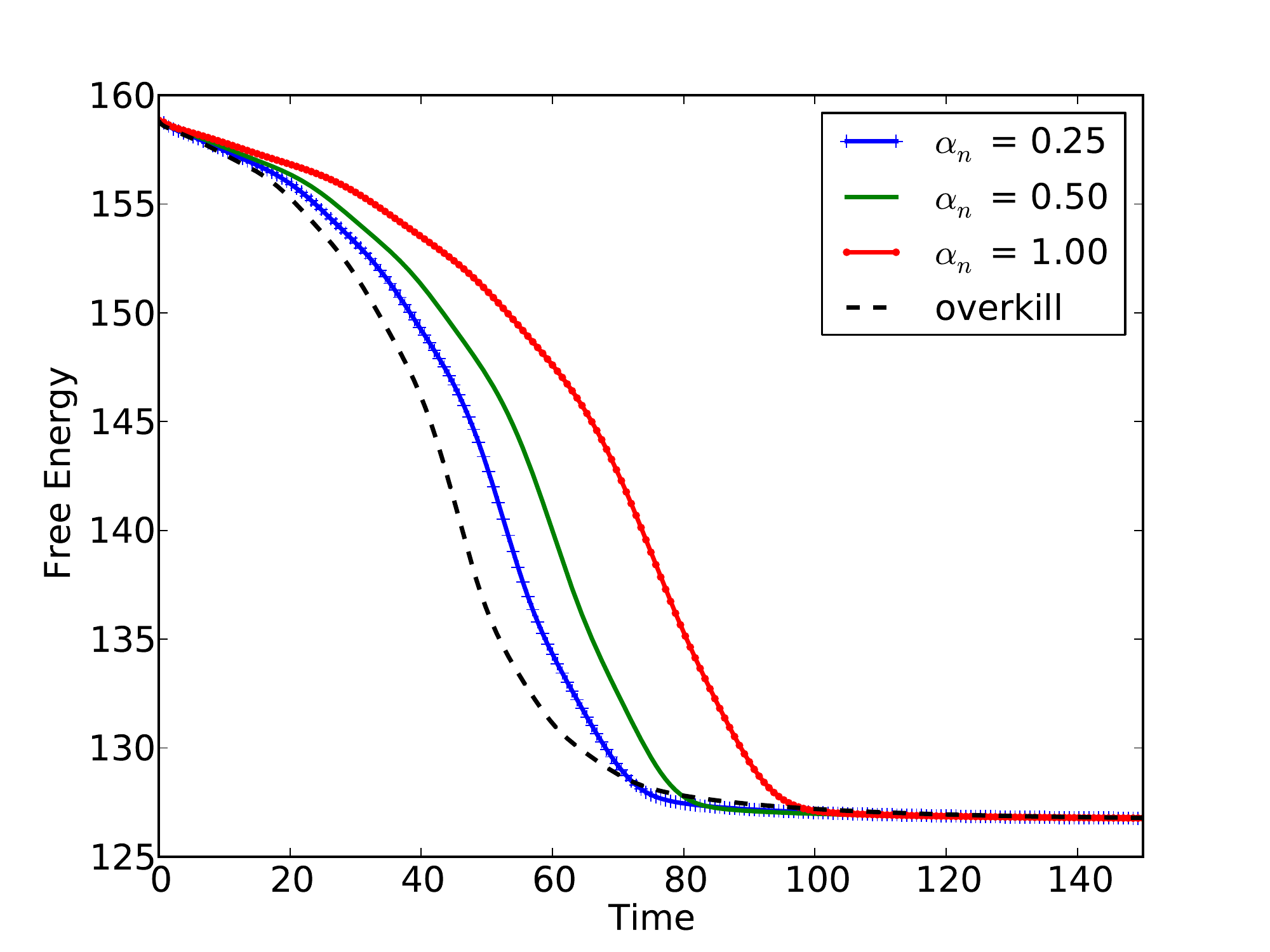}}\label{pfc2d_dt0_75}~\!~
{\includegraphics[trim=0cm 0.3cm 1.4cm 1.3cm, clip=true,width=0.48\columnwidth]{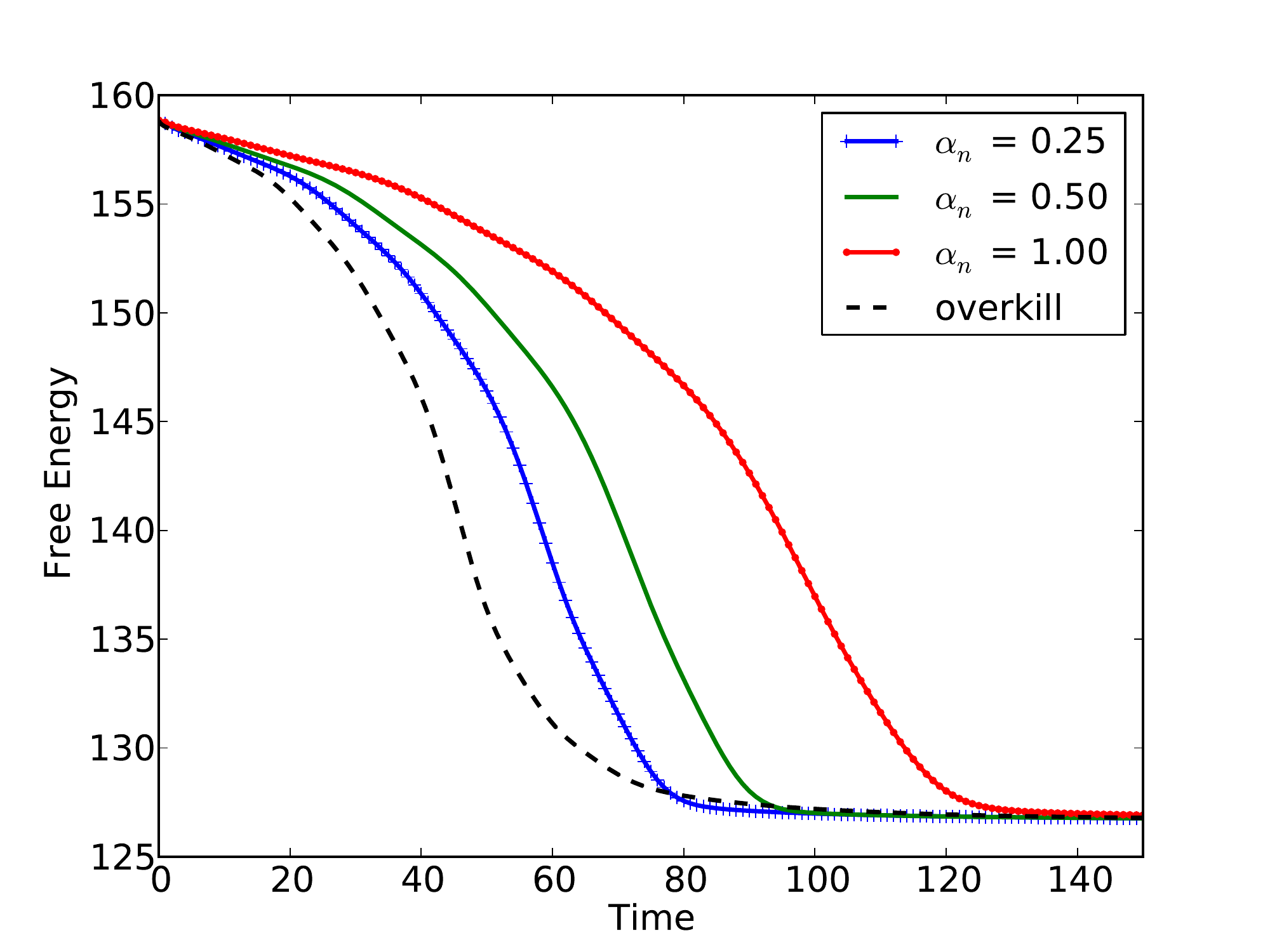}}\label{pfc2d_dt1} \\
\qquad \qquad(c) $\Delta t=0.75$ \qquad \qquad \qquad \qquad \qquad (d) $\Delta t=1.0$\qquad \qquad \\
\end{tabular}
\caption{Stabilization parameter variation in two dimensions. The free energy is plotted as a function of time using time step sizes (a)~$\Delta t=0.25$, (b)~$\Delta t=0.5$, (c)~$\Delta t=0.75$ and (d)~$\Delta t=1.0$, respectively. Increasing the stabilization parameter $\alpha_n$ or the time step size $\Delta t$ results in a less accurate dynamical representation, but converges to the correct steady state solution. The mesh consists of $\left[128 \times 133\right]$ quadratic $\mathcal{C}^0$ elements, such that the spatial error could be considered negligible in the simulations.}\label{fig:pfc2d_alpha_variation}
\end{figure}

%
We proceed to study the temporal order of accuracy of the method. Using the same spatial resolution as our reference solution, we perform simulations over a range of time step sizes, and focus on the $\mathcal{L}^2$-error norm in space
\begin{align*}
||e||_{2}=\left(\int_\Omega\left(\phi^h-\phi^h_*\right)^2d\mathbf{V}\right)^{^{1}/_{2}}
\end{align*}
where $\phi^h$ are the coarse-in-time solutions and $\phi^h_*$ corresponds to the reference solution. We compute this error at $t=150$, point in time at which the crystal lattice has already grown over the whole domain.
The convergence in time is shown in Fig.~\ref{fig:convergence}, where we observe that the numerical scheme is indeed second-order accurate. 
%
The maximum relative error in mass with different time step sizes is shown in Fig.~\ref{fig:conservation}. We conclude that the mass is indeed conserved in all cases, as the maximum relative error in mass for different time step sizes stays below $10^{-9}$.
Fig.~\ref{fig:energy} shows the time evolution of free energy with different time step sizes. Free-energy monotonicity is verified for all the cases, as no increases in free energy are observed. The increase in time step size nonetheless leads to a poorer dynamical representation of the free energy evolution, which is consistent with other published results~\cite{Zhang3,Gomez}. Care has to be taken when choosing $\alpha_n$, as increasing the stabilization parameter has a negative effect on the free energy approximation. This can be seen in Figs.~\ref{fig:pfc2d_alpha_variation}(a), \ref{fig:pfc2d_alpha_variation}(b), \ref{fig:pfc2d_alpha_variation}(c) and \ref{fig:pfc2d_alpha_variation}(d), where free energy is plotted for different values of $\alpha_n$ and $\Delta t$. Nonetetheless, as long as the stabilization parameter $\alpha_n$ complies with the bound presented in equation~$(29)$, free energy is dissipated. Even though the dynamics of the equation are influenced by the time step size, the method converges to the right steady state solution. This could be an advantage if what is looked for is the steady state solution to a problem, such as in control of dynamical systems~\cite{Levine}. The use of $\alpha_n$ slows down the dynamics of the equation, and an effective time-step size needs to be determined. We plan to study this point further in future work~\cite{Vignal1}.

The results in this section validate numerically the theoretical results presented in section~\ref{subset:time} with regards to this numerical formulation, and prove that it is indeed mass conserving, unconditionally energy-stable, and is second-order accurate in time.

\subsection{Three dimensional simulations: Crystalline growth in a supercooled liquid}
In this section, we deal with the three dimensional version of the example described in section~\ref{sec:validation}, as well as a more challenging case, where two crystallites oriented in different directions are grown in the same domain. The PFC equation in this latter case is able to capture the emergence of grain boundaries.
\subsubsection{Crystalline growth in a supercooled liquid}\label{sec:1crystal}
\begin{figure}
\centering
\begin{tabular}{cc}
 \includegraphics[trim=1.1cm 3.9cm 0cm 0.75cm, clip=true,width=0.95\textwidth]{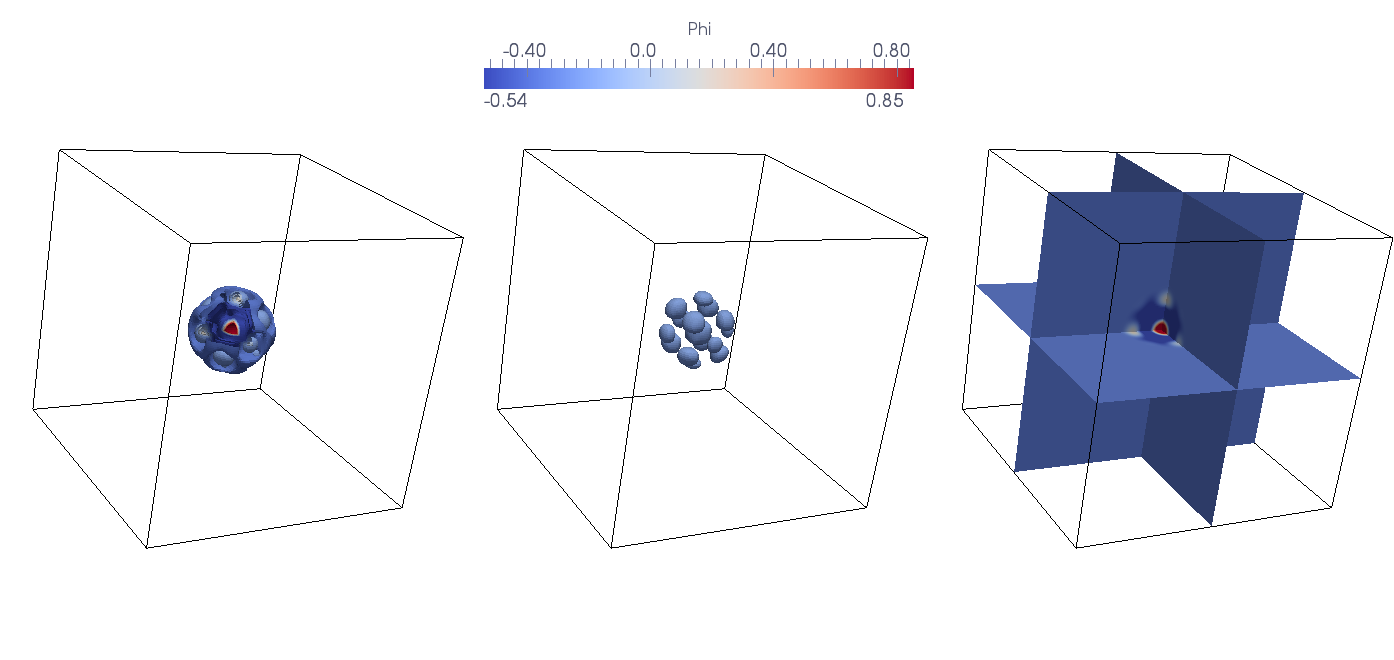}\label{pfc3d0}\\
(a) $t=0$\\
{\includegraphics[trim=1.1cm 3.9cm 0cm 0.75cm, clip=true,width=0.95\columnwidth]{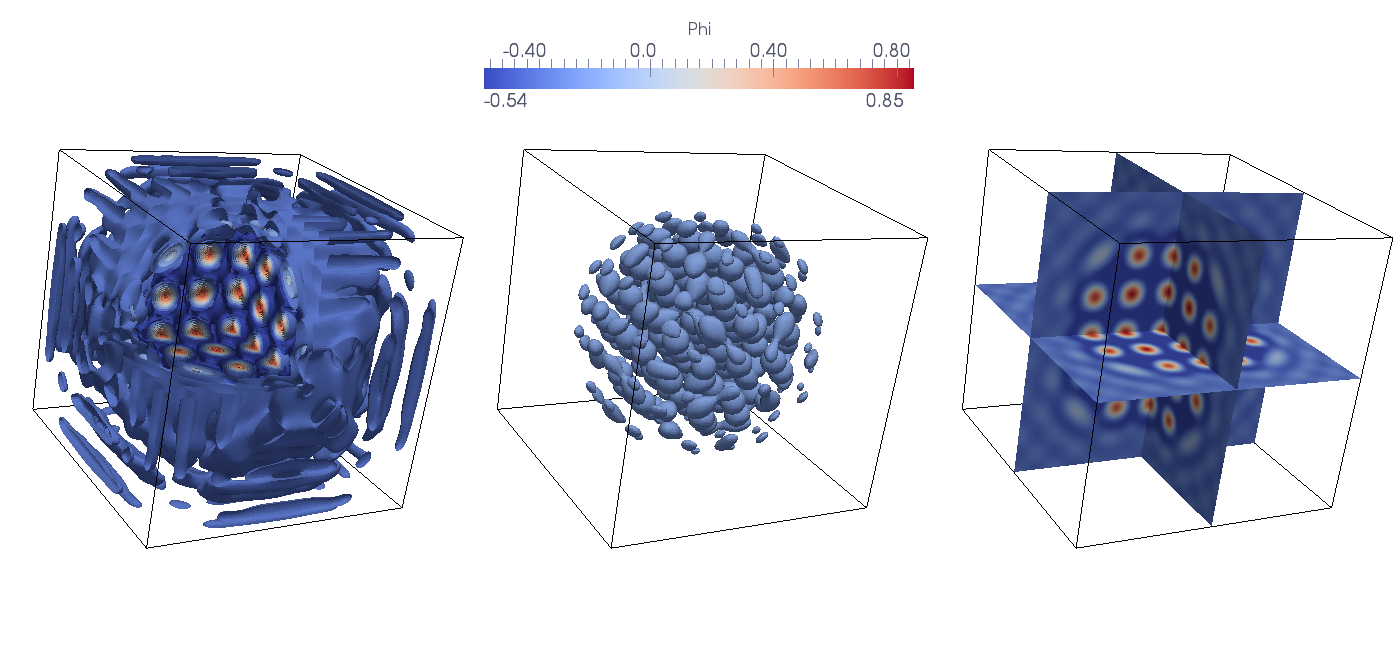}\label{pfc3d100}} \\
 (b) $t=100$\\
 {\includegraphics[trim=1.1cm 3.9cm 0cm 0.75cm, clip=true,width=0.95\columnwidth]{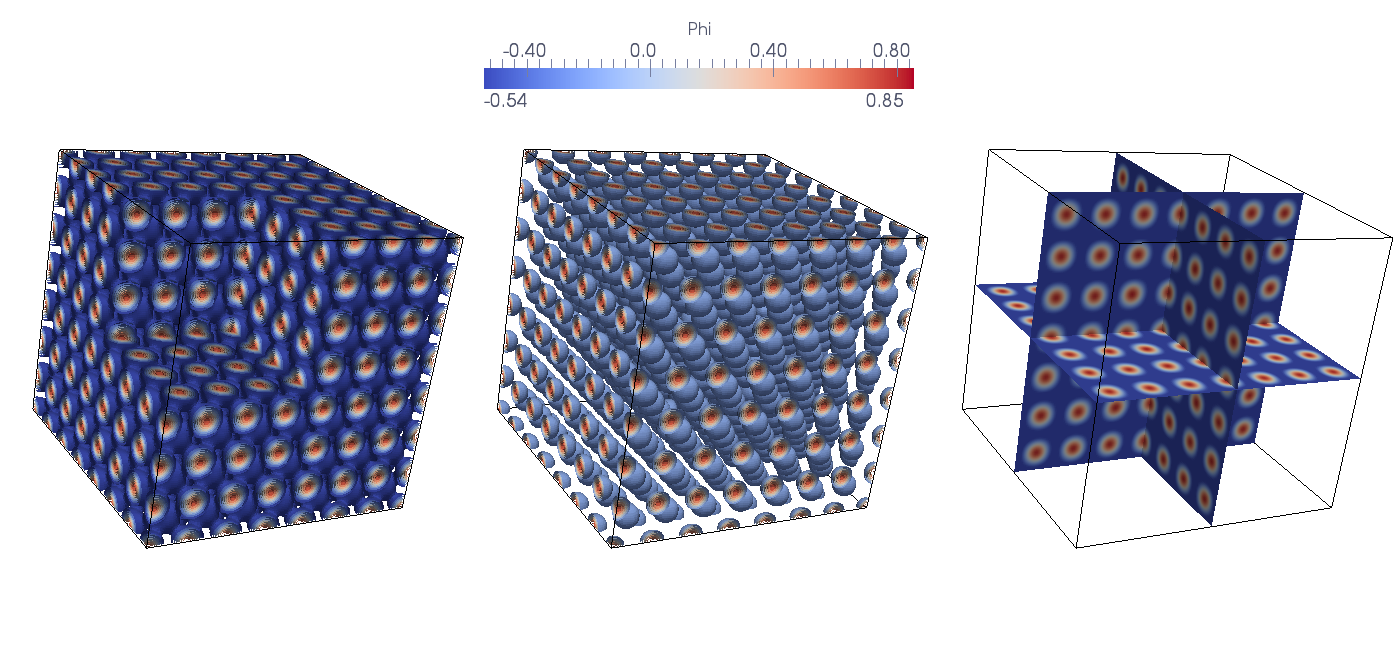}\label{pfc3d250}}  \\
(c) $t=250$ \ (\text{steady state}) \\
\end{tabular}
\caption{Crystal growth in a supercooled liquid in three dimensions. The images show the evolution of one crystallite surrounded by liquid. The labels indicate the computational time. On the left-hand side, we show isosurfaces of the solution, in the middle we present the same isosurfaces where a thresholding filter has been applied to only show the atoms, such that the periodic nature of the lattice is clear, while on the right-hand side we present slices of the solution across the indicated planes.}\label{fig:pfc3d}
\label{fig:snapshotssingle}
\end{figure}

\begin{figure}
\begin{center}
\begin{tikzpicture}
  \node (img1) {\includegraphics[width=1.0\textwidth]{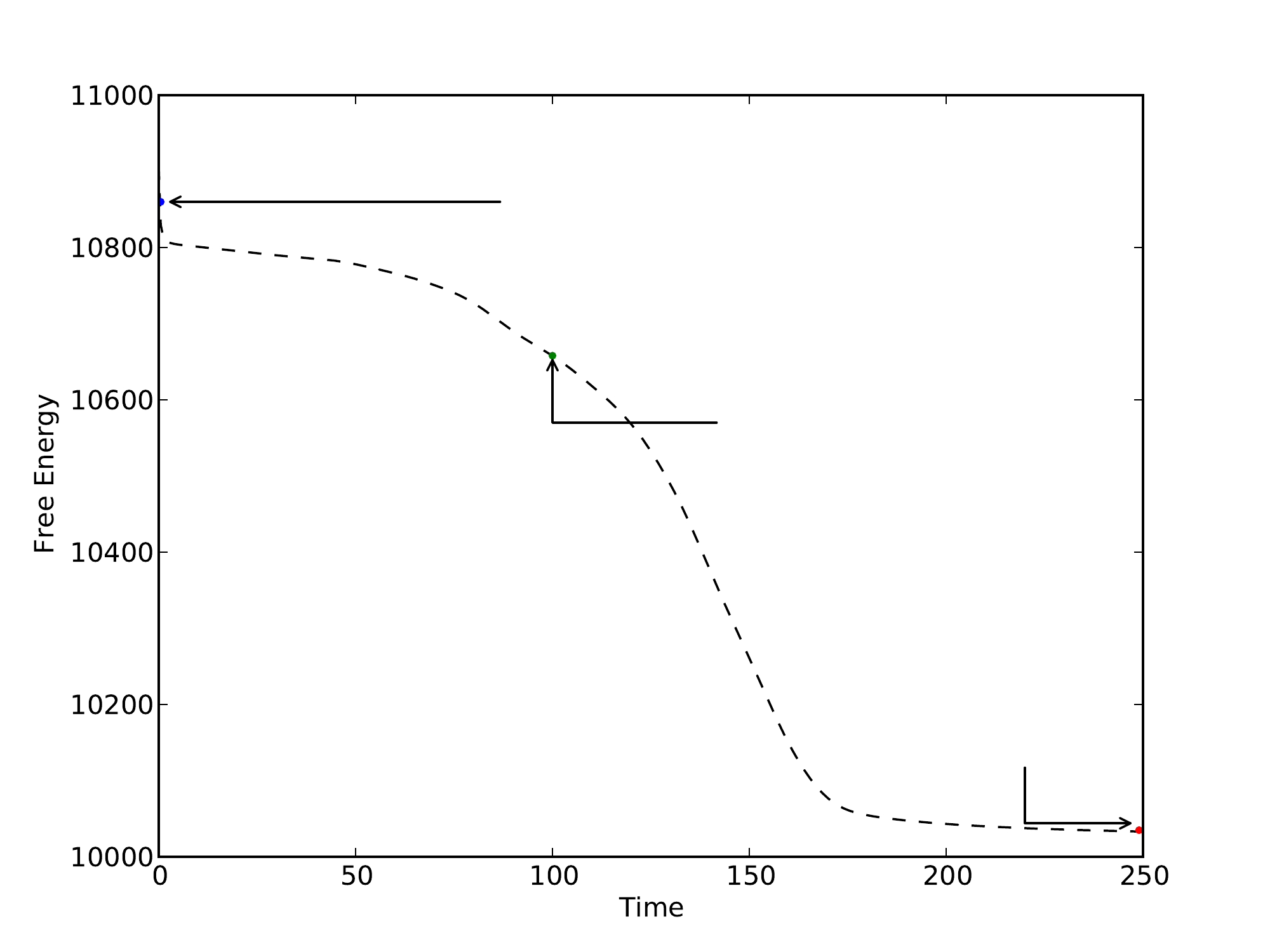}};
  \node (img2) at (img1.center)[yshift = 2.75cm][xshift=+0.2cm] {\includegraphics[trim=0cm 0cm 0cm 0.1cm, clip=true,width=0.2\textwidth]{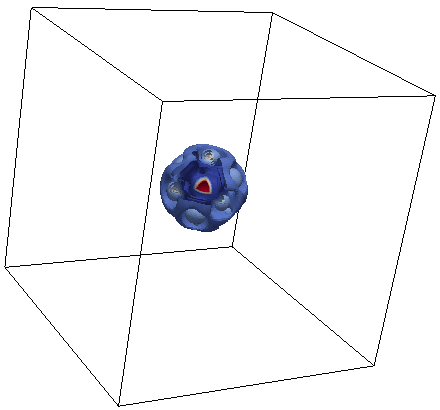}};
  \node (img3) at (img1.center)[yshift = 0.4cm][xshift=2.4cm] {\includegraphics[trim=0cm 0cm 0cm 0.1cm, clip=true, width=0.2\textwidth]{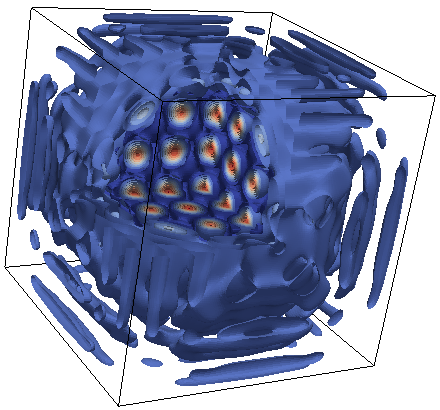}};
  \node (img3) at (img1.center)[yshift = -2.1cm][xshift=4.1cm] {\includegraphics[trim=0cm 0cm 0cm 0.1cm, clip=true, width=0.2\textwidth]{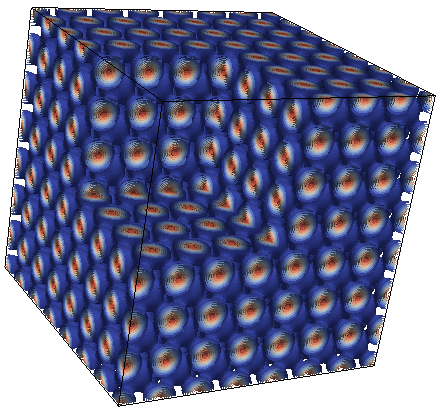}};
\end{tikzpicture}
\caption{Free energy evolution of a single crystal. The free energy is monotonically decreasing while the mass remains constant throughout the simulation (the maximum relative error stays below $10^{-9}$), which was run using a mesh composed of $[150]^3$ linear elements. A time step size of $0.5$ was used, with an $\alpha_n$ value of~0.5.}\label{fig:fe_evolution_3d1}
\end{center}
\end{figure}

In this example, the growth of a single crystal with a BCC structure is simulated. Mathematically, the crystallite is now defined as~\cite{Elder2010,Jaatinenext}
\begin{align}\label{eq:BCC_crystallite}
\phi_{\text{BCC}}\left(\mathbf{x}\right)&=\text{cos}\left(xq_{\text{BCC}}\right)\text{cos}\left(yq_{\text{BCC}}\right)+\text{cos}\left(xq_{\text{BCC}}\right)\text{cos}\left(zq_{\text{BCC}}\right)\nonumber\\
&+\text{cos}\left(yq_{\text{BCC}}\right)\text{cos}\left(zq_{\text{BCC}}\right),
\end{align}
where $x$, $y$ and $z$ represent the three-dimensional Cartesian coordinates and $q_{BCC}$ represents a wavelength related to the BCC crystalline structure. The computational domain is $\Omega = [0,20\pi]^3$, with periodic boundary conditions being assumed again in all directions. Similarly to what is done in equation~\eqref{eq:initial} for the two-dimensional case, the initial condition is defined as
\begin{align}
\phi_o\left(\mathbf{x}\right)=\bar{\phi}_{BCC}+\omega\left(\mathbf{x}\right)\left(A\phi_{BCC}\left(\mathbf{x}\right)\right),
\end{align}
where $\bar{\phi}_{BCC}$ represents again the average density of the liquid domain, and $A$ represents an amplitude of the fluctuations in density. To ensure the stability of the BCC phase, the parameters of the equation are given the following values
\begin{align*}
\epsilon=0.35, \quad \bar{\phi}_{BCC}=-0.35, \quad q_{BCC}=\dfrac{{1}}{\sqrt{2}}, \quad A=1.
\end{align*}
The initial crystallite is placed in the centre of the domain. Similarly to what happens in the two dimensional case, the crystal grows at the expense of the liquid. Snapshots of the solution can be observed in Fig.~\ref{fig:snapshotssingle}. The figure shows that the initial BCC pattern is repeated over the whole domain, until reaching a steady state. The simulation uses a uniform grid composed of $[150]^3$ linear elements, and a time step size $\Delta t=0.5$. The stabilization parameter $\alpha_n$ is set to $0.5$. The free energy evolution for the simulation is shown in Fig.~\ref{fig:fe_evolution_3d1}. There are no increases in free energy. The mass also remains constant throughout the simulation.
%

\subsubsection{Polycrystalline growth of BCC crystals}
\begin{landscape}
\begin{figure}
\centering
\begin{tabular}{cc}
{\includegraphics[width=0.5\columnwidth]{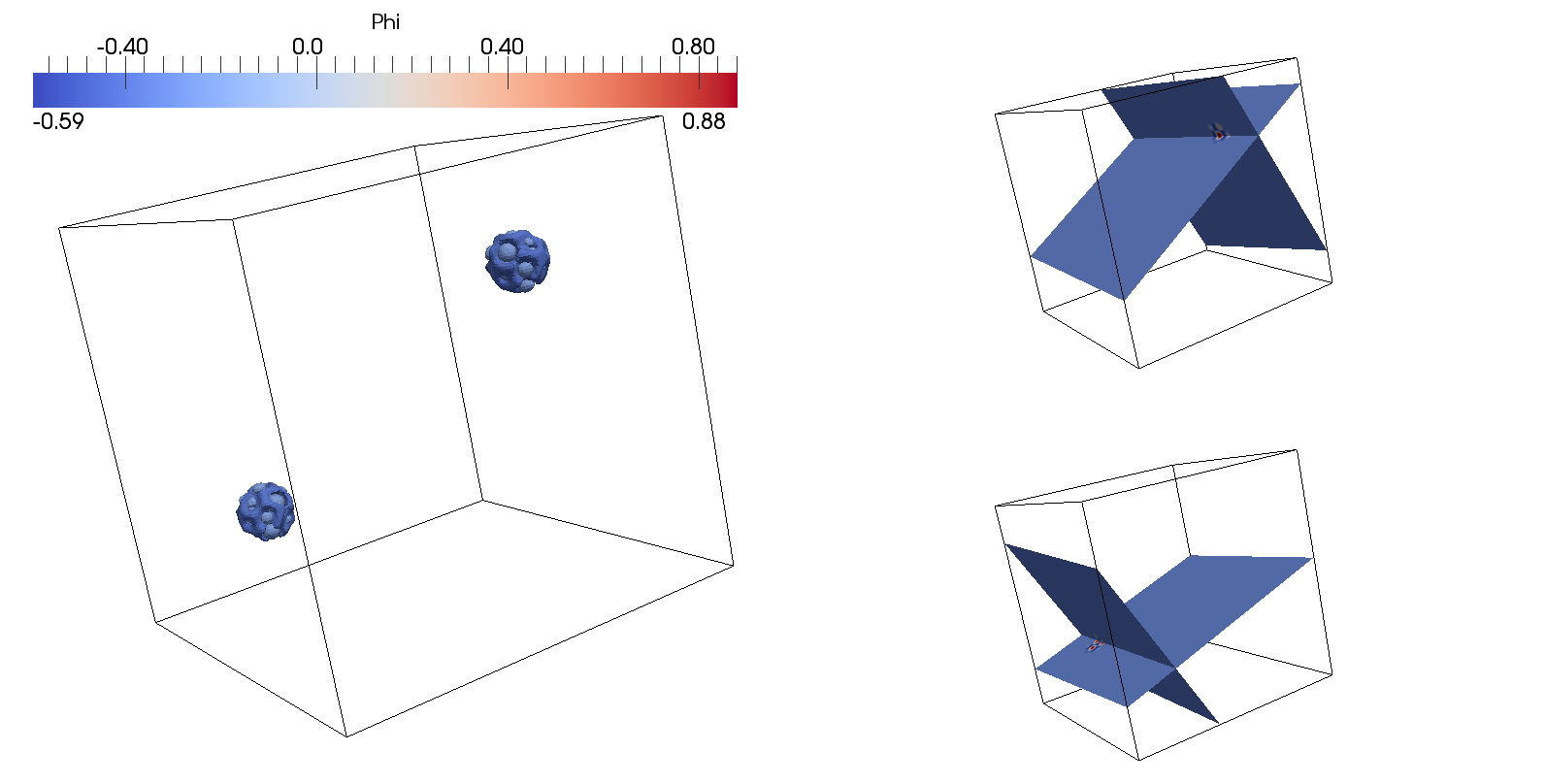}}  & {\includegraphics[width=0.5\columnwidth]{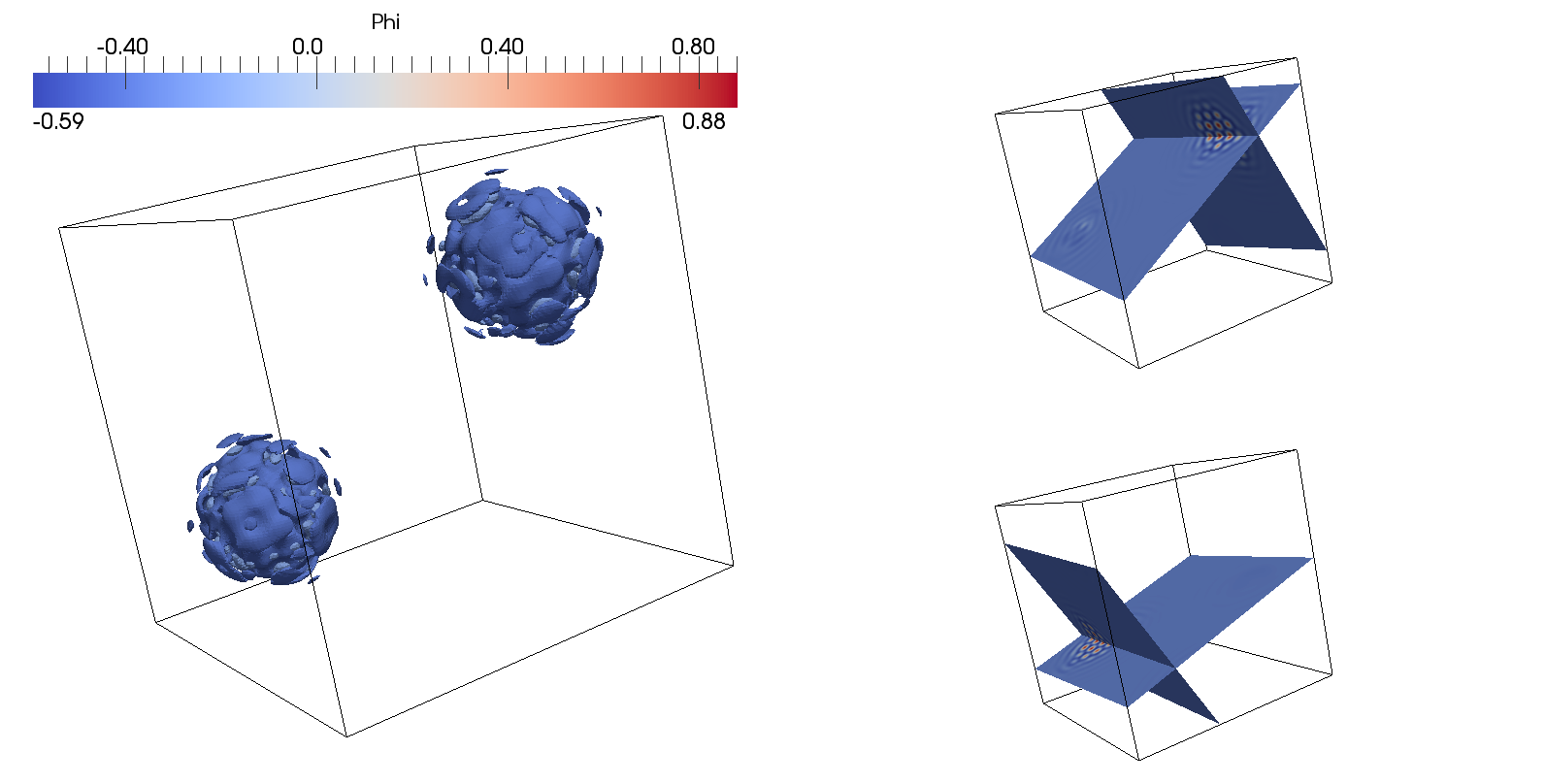}} \\
(a) t = 0. & (b) t = 100. \\
{\includegraphics[width=0.5\columnwidth]{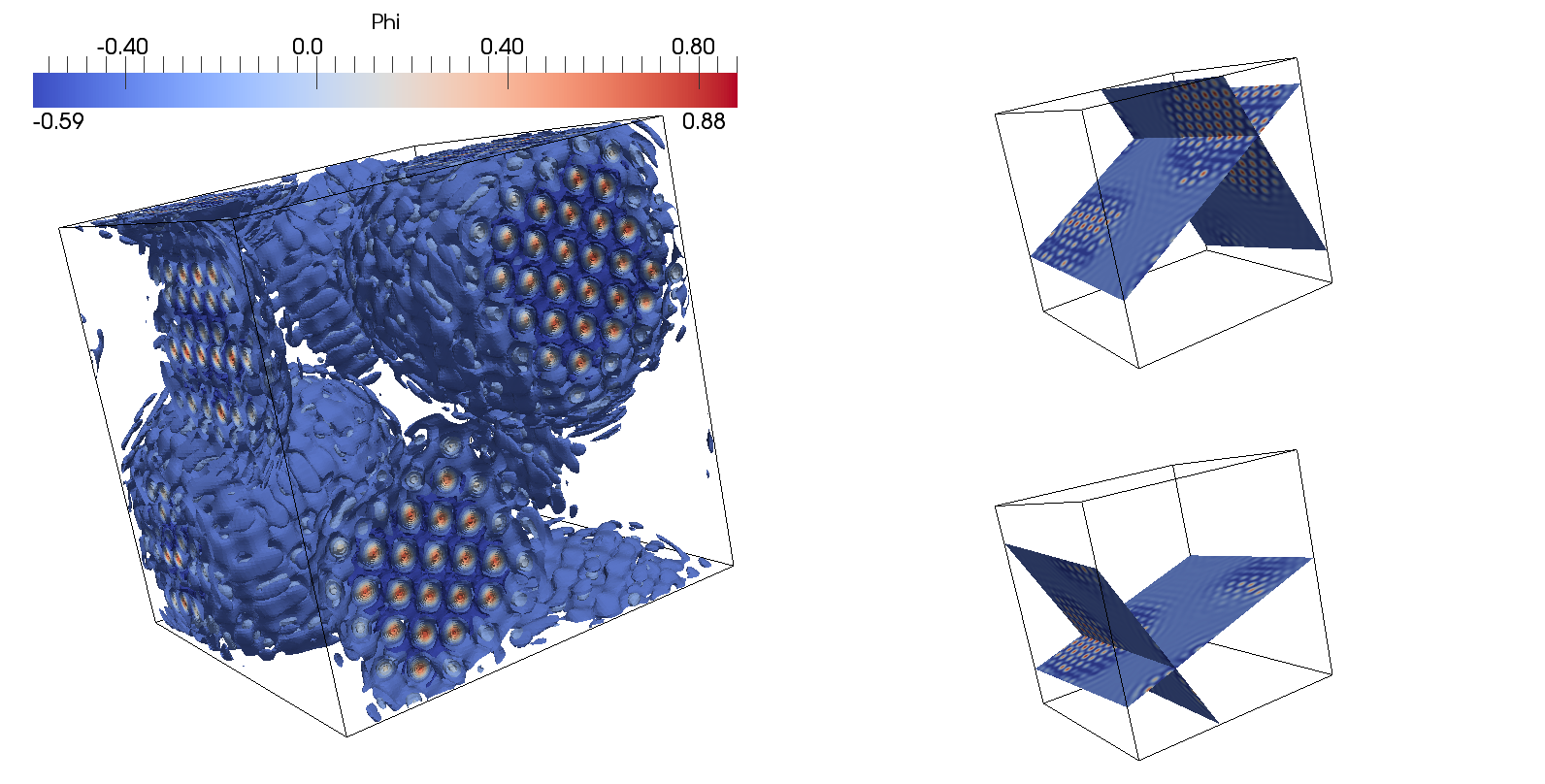}} & {\includegraphics[width=0.5\columnwidth]{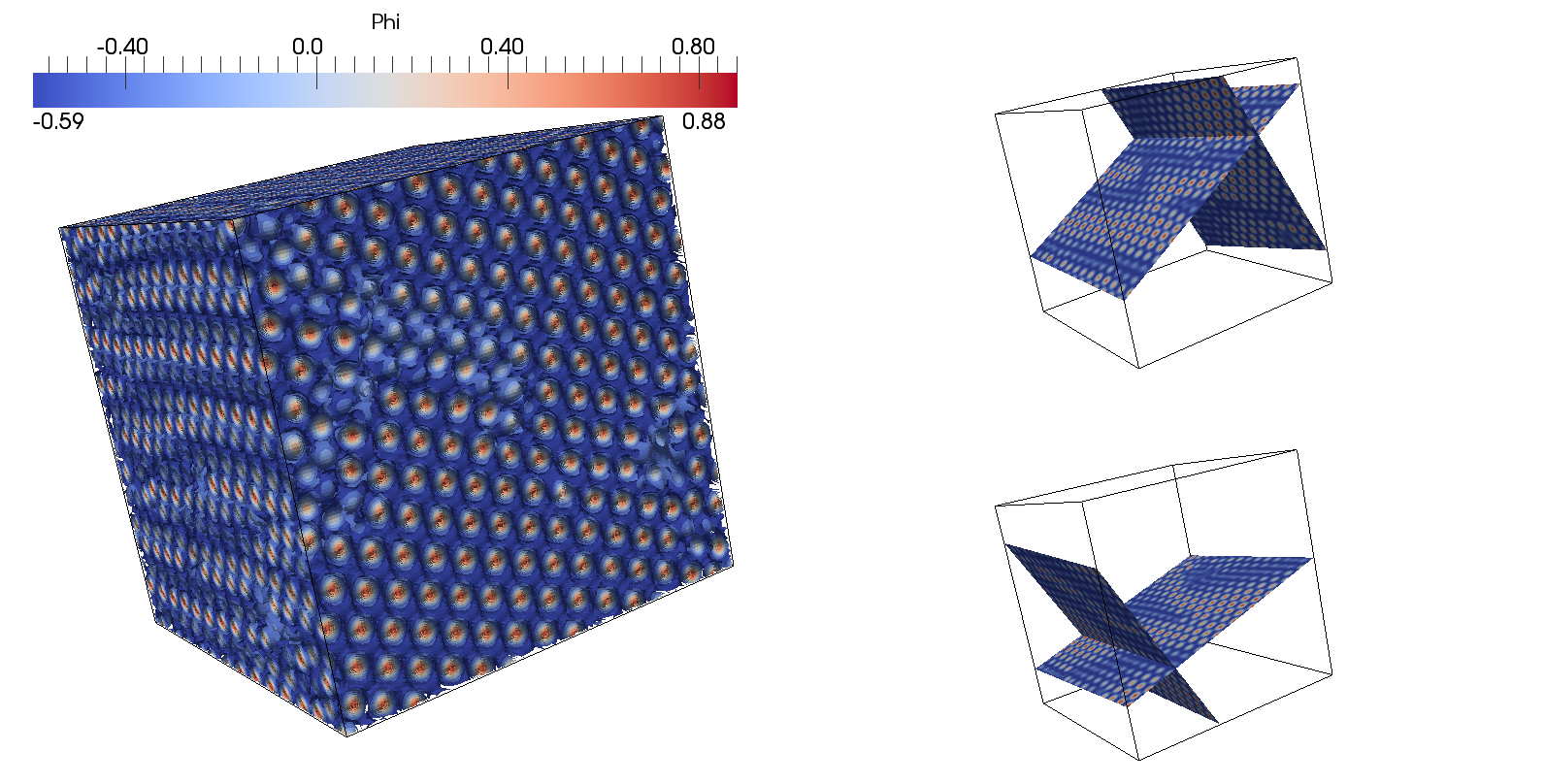}}  \\
(c) t = 250. & (d) t = 500. \\
\end{tabular}
\caption{Polycrystalline growth in a supercooled liquid in three dimensions. The images show isocontours of the atomistic density field, where two crystallites are initially placed in a domain with different orientations. Grain boundaries emerge once the crystals meet. The labels indicate the computational time, while the mesh used $[150]^3$ linear elements. A time step size of $0.5$ was used, with an $\alpha_n$ value of~0.5.}
\label{fig:policrystal}\end{figure}
\end{landscape}

\begin{figure}
\begin{center}
\begin{tikzpicture}
  \node (img1) {\includegraphics[width=1.0\textwidth]{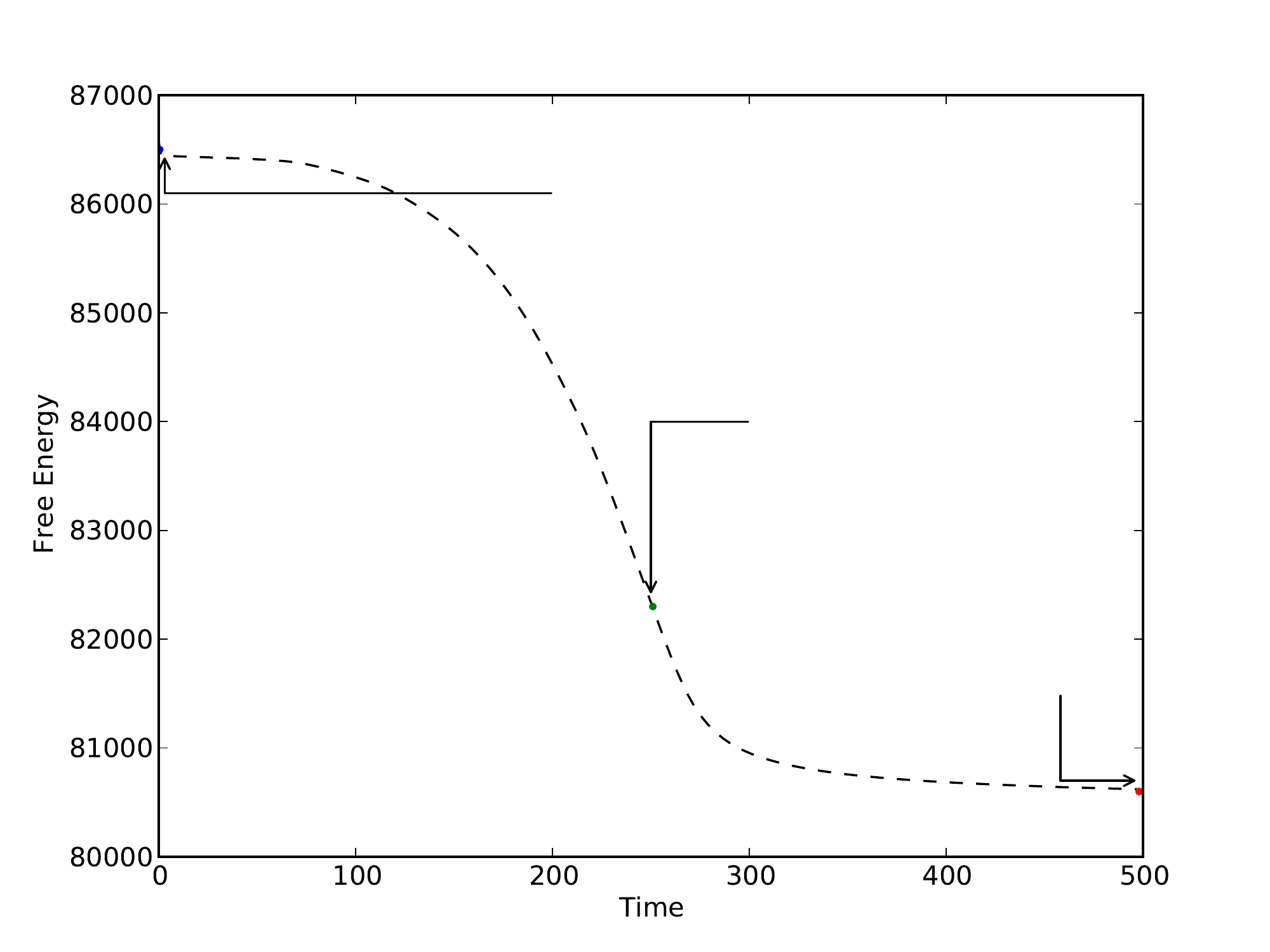}};
  \node (img2) at (img1.center)[yshift = 2.8cm][xshift=0.5cm] {\includegraphics[trim=3.6cm 1.5cm 2.6cm 4cm, clip=true,width=0.19\textwidth]{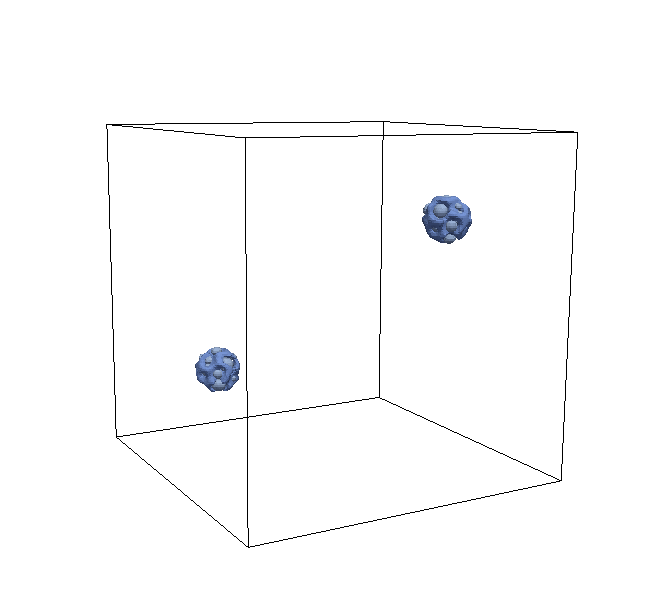}};
  \node (img3) at (img1.center)[yshift = 0.5cm][xshift=2cm] {\includegraphics[trim=3.6cm 1.5cm 2.6cm 4cm, clip=true,width=0.19\textwidth]{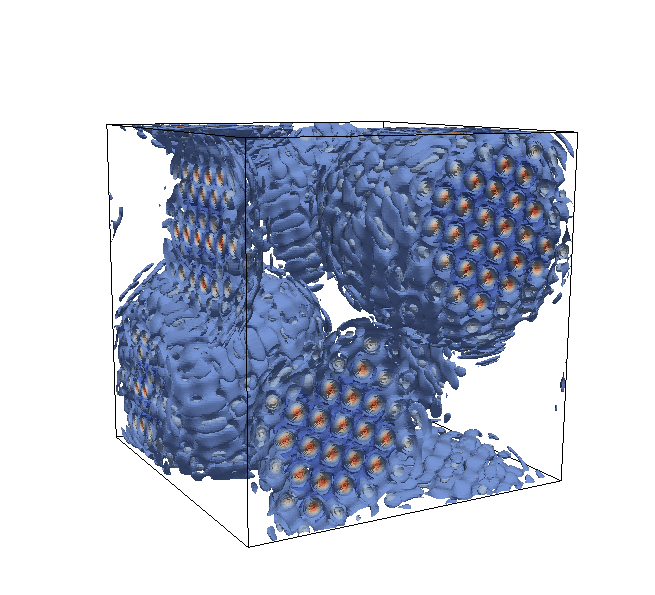}};
  \node (img3) at (img1.center)[yshift = -1.7cm][xshift=4cm] {\includegraphics[trim=3.6cm 1.5cm 2.6cm 4cm, clip=true,width=0.19\textwidth]{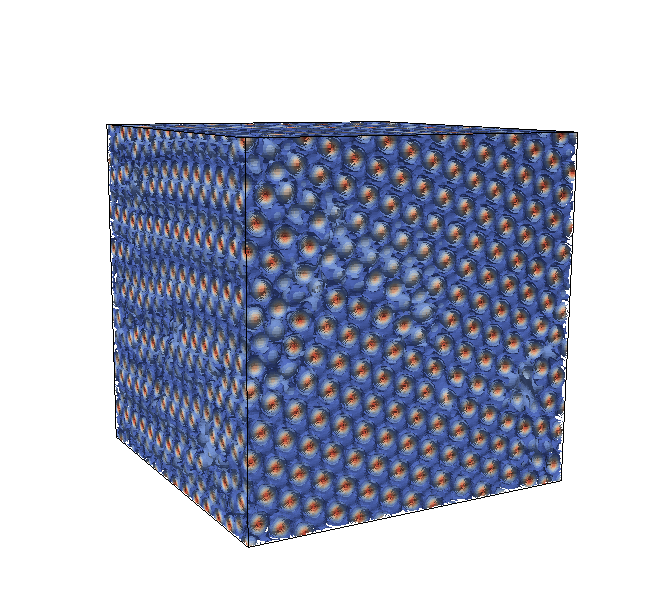}};
\end{tikzpicture}
\caption{Free energy evolution of two crystals. The free energy is monotonically decreasing while the mass remains constant throughout the simulation (the maximum relative error stays below $10^{-9}$), which was run using a mesh composed of $150^3 \mathcal{C}^0$ linear elements. A time step size of $0.5$ was used, with an $\alpha_n$ value of~0.5.}\label{fig:fe_evolution_3d}
\end{center}
\end{figure}

As a more challenging example, we present a case of polycrystalline growth, where two initial crystallites with a BCC configuration oriented in different directions are placed in the domain. They are set at different angles, so as to eventually observe the emergence of grain boundaries when both crystallites meet. The computational domain is $\Omega=\left[0,40\pi\right]^3$, and periodic boundary conditions are imposed in all directions. A uniform mesh comprised of $[150]^3$ linear elements is used, along with a time step size $\Delta t=0.5$. The stabilization parameter $\alpha_n$ is set to $0.5$. 
A system of local Cartesian coordinates $(x_C,y_C,z_C)$ was used to generate the crystallites in different directions, by doing an affine transformation of the global coordinates $(x,y,z)$ to produce a rotation $\beta$ along the z axis, with
\begin{align}\left(
    \begin{array}{c}
      x_C \\
      y_C\\
      z_C 
    \end{array}
  \right) = \left(\begin{array}{c}
      \text{cos}\left(\beta\right)x - \text{sin}\left(\beta\right)y \\
       \text{cos}\left(\beta\right)x + \text{sin}\left(\beta\right)y \\
      z 
    \end{array}
  \right).
\end{align}
The first crystallite was defined as in equation~\eqref{eq:BCC_crystallite}, with $\beta=0$, while the second one was rotated by an angle $\beta=\dfrac{\pi}{8}$. The same equation parameters that were used in section~\ref{sec:1crystal} are used in this example, and result in the simulation shown in Fig~\ref{fig:policrystal}. Grain boundaries appear when the two crystals meet while growing, given the orientation mismatch. The free energy evolution is shown in Fig.~\ref{fig:fe_evolution_3d}, where no increases in free energy are seen. 

\begin{figure}
\centering
 {\includegraphics[trim=0cm 0.2cm 1.5cm 1.cm, clip=true,width=1\textwidth]{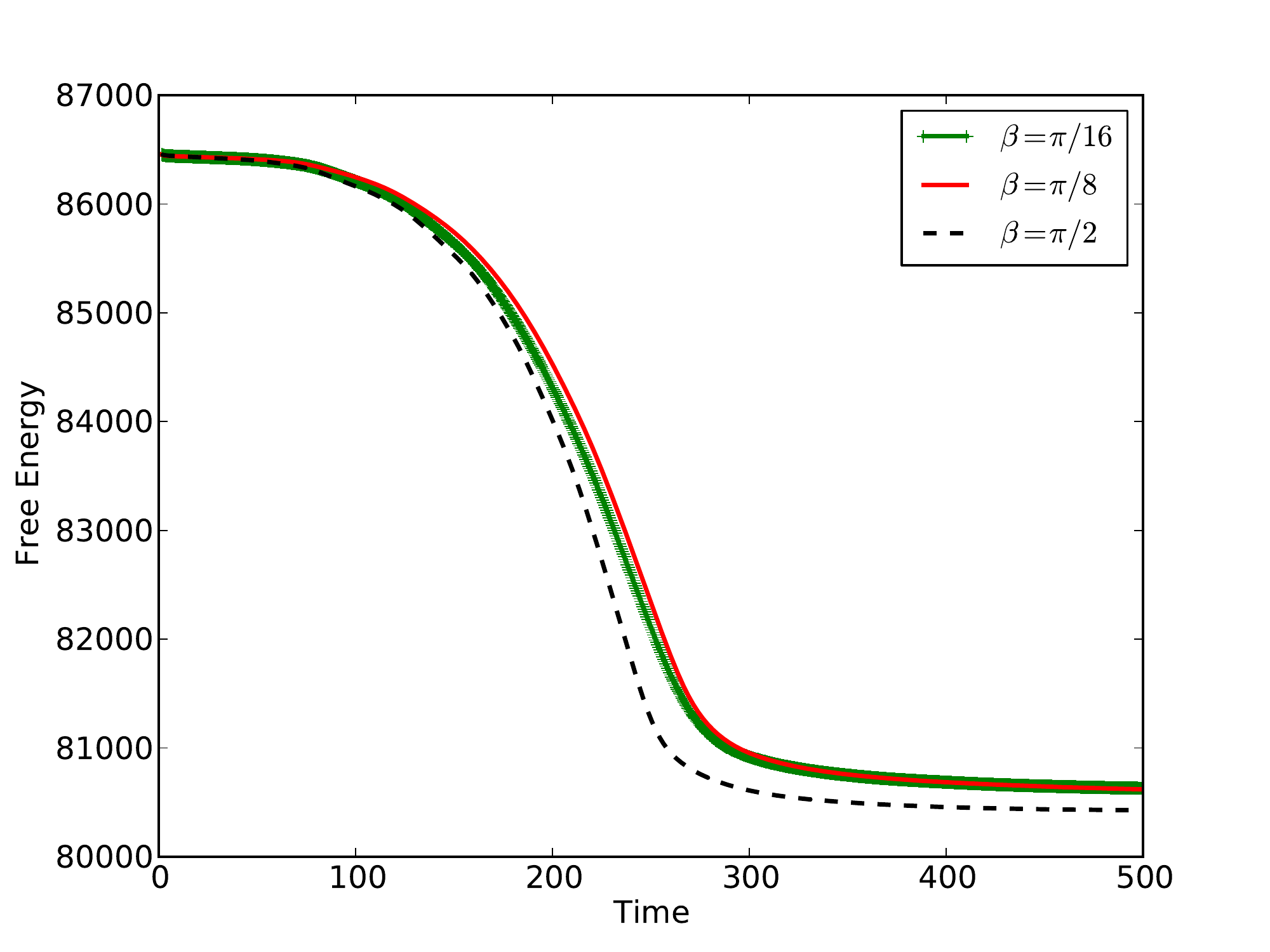}\label{pfc2d_dt0_2}}\\
\caption{Effect of rotation angle on the crystallites. The plotted solutions use a time step size $\Delta t = 0.5$, a stabilization parameter $\alpha_n=0.5$, and $[150]^3$~linear elements. Three different rotation angles $\beta$ are considered.}\label{fig:pfc3d_rotation}
\end{figure}

Changing the rotation angle $\beta$ can have an effect on the free energy of the system, as it influences the grain boundary that is formed. The free energy evolution is plotted in Fig.~\ref{fig:pfc3d_rotation} for three different values of $\beta$. In the two cases where the rotation angle is relatively small \Big(i.e, $\beta=\dfrac{\pi}{8}$ and $\beta=\dfrac{\pi}{16}$\Big), the same steady state is reached, as the equation leads both systems to the same energetically minimal state. On the other hand, when the change in $\beta$ is larger \Big($\dfrac{\pi}{2}$\Big), the free energy value at steady state differs significantly. The grain boundary formed is considerably different than the ones considered before, as the two grains meet at a significantly different position. Further studies are needed to conclude if the free energy differences are qualitatively accurate and compare well with experiments.

\section{Conclusion}\label{s:conc}In this work we present a provably, unconditionally stable algorithm to solve the phase-field crystal equation. This algorithm conserves mass, possesses strong energy stability and is second-order accurate in time. Theoretical proofs are presented, along with numerical results that corroborate them. The numerical formulation recurs to a mixed finite element formulation that deals with a system of three coupled, second-order equations. Three dimensional results involving polycrystalline growth are also presented, showcasing the robustness of the method. The implementation was done using PetIGA, a high performance isogeometric analysis framework, and the codes are freely available to download~\footnote{https://bitbucket.org/dalcinl/petiga}.

\section{Appendix}\label{s:appendix}
\subsection{Jacobian for the $2+2+2$ mixed form}\label{jacobian222}
The Jacobian components of $\mathbf{K}$ in equation~(\ref{jacobian_components222}) are defined as
\begin{align}
{K}_{AB}^{\phi\phi}&=\left(N_A,\dfrac{N_B}{\Delta t}\right)_{\Omega},                                                                                     \label{jac22200} \\
{K}_{AB}^{\phi\sigma}&=\left(\nabla N_A,\nabla N_B\right)_{\Omega},                                                                                    \label{jac22201} \\
{K}_{AB}^{\phi\theta}&=0,                                                                                                                                                          \label{jac22202} \\
{K}_{AB}^{\sigma\phi}&=\dfrac{1}{2}\left(N_A,N_B\right)_{\Omega}\left(\Psi_c'''\left(\phi_{n+1}^h\right)\jump{\phi_n^h}-\Psi_c''\left(\phi_{n+1}^h\right)+\Psi_e''\left(\phi_{n}^h\right)\right) \nonumber\\
&-\alpha_n\Delta t\left(\nabla N_A, \nabla N_B\right)_{\Omega},                                                                                              \label{jac22210}\\
{K}_{AB}^{\sigma\sigma}&=\left(N_A,N_B\right)_{\Omega},                                                                                                       \label{jac22211}\\
{K}_{AB}^{\sigma\theta}&=-\left(N_A,N_B\right)_{\Omega}+\left(\nabla N_A,\nabla N_B\right)_{\Omega},                               \label{jac22212}\\
{K}_{AB}^{\theta\phi}&=-\dfrac{1}{2}\left(N_A,N_B\right)_{\Omega}+\dfrac{1}{2}\left(\nabla N_A,\nabla N_B\right)_{\Omega}.\label{jac22220}\\
{K}_{AB}^{\theta\sigma}&=0,                                                                                                                                                       \label{jac22221} \\
{K}_{AB}^{\theta\theta}&=\left(N_A,N_B\right)_{\Omega}.                                                                                                          \label{jac22222}
\end{align}
\subsection{Running the code}\label{codes}
Both PETSc and PetIGA are regularly maintained and updated, so it is worthwhile to download their respective repositories through the version control systems Git\footnote{\url{http://git-scm.com}} and Mercurial\footnote{\url{http://mercurial.selenic.com/}}. These tools can be used to clone the PETSc and PetIGA repositories with the following commands 
\begin{itemize}
\item \texttt{git clone https://bitbucket.org/petsc/petsc}
\item \texttt{hg clone https://bitbucket.org/dalcinl/PetIGA}
\end{itemize}
PETSc must be configured and installed before installing PetIGA. After completing the PetIGA installation, the igakit repository can be cloned.
\begin{itemize}
\item \texttt{hg clone https://bitbucket.org/dalcinl/igakit}
\end{itemize}
Igakit is a Python-based pre-processing and post-processing tool for PetIGA. Further information on these software packages can be found in~\cite{petsc1,petsc2,framework,Collier,petiga}, and the discretization proposed in this work can be found in the \texttt{demo/} directory of the PetIGA sources.

\section*{Acknowledgements}
We would like to acknowledge the open source software packages that
made this work possible: PETSc~\cite{petsc1,petsc2}, NumPy~\cite{numpy}, matplotlib~\cite{matplotlib}, {ParaView}~\cite{paraview}. 

This work was supported by the Center for Numerical Porous Media~(NumPor) at King Abdullah University of Science and Technology (KAUST).
\section{References}
\bibliographystyle{unsrt}	
\bibliography{PFC1}	
\end{document}